\documentclass[review]{elsarticle}

\usepackage{lineno,hyperref}
\modulolinenumbers[5]

\journal{arXiv}

\usepackage{amsmath, amssymb, amsthm}
\newtheorem{theorem}{Theorem}
\newtheorem{proposition}{Proposition}

\usepackage{fixltx2e}
\usepackage[usenames]{color}
\usepackage{graphicx}
\usepackage{tikz}

\ifodd 0
\newcommand{\rev}[1]{{\color{blue}#1}} %revise of the text
\newcommand{\com}[1]{\textbf{\color{red}(COMMENT: #1)}} %comment of the text
\newcommand{\edt}[1]{\textbf{\color{magenta}#1}} %edit this text
\newcommand{\clar}[1]{\textbf{\color{green}(NEED CLARIFICATION: #1)}}

\else
\newcommand{\rev}[1]{#1}
\newcommand{\com}[1]{}
\newcommand{\edt}[1]{}
\newcommand{\clar}[1]{}
\fi
\begin{document}

\begin{frontmatter}

\title{A Tale of Two Mechanisms:
Incentivizing Investments in Security Games}

\author{Parinaz Naghizadeh and Mingyan Liu\fnref{myfootnote}}
\address{Department of EECS, University of Michigan, Ann Arbor}
\fntext[myfootnote]{\{naghizad, mingyan\}@umich.edu}

\begin{abstract}
In a system of interdependent users, the security of an entity is affected not only by that user's investment in security measures, but also by the positive externality of the security decisions of (some of) the other users. The provision of security in such system is therefore modeled as a public good provision problem, and is referred to as a security game. In this paper, we compare two well-known incentive mechanisms in this context for incentivizing optimal security investments among users, namely the Pivotal and the Externality mechanisms. The taxes in a Pivotal mechanism are designed to ensure users' voluntary participation, while those in an Externality mechanism are devised to maintain a balanced budget. We first show the more general result that, due to the non-excludable nature of security, no mechanism can incentivize the socially optimal investment profile, while at the same time ensuring voluntary participation and maintaining a balanced budget for all instances of security games. To further illustrate, we apply the Pivotal and Externality mechanisms to the special case of \emph{weighted total effort} interdependence models, and identify some of the effects of varying interdependency between users on the budget deficit in the Pivotal mechanism, as well as on the participation incentives in the Externality mechanism. 
\end{abstract}


\end{frontmatter}

%\linenumbers

%%%%%%%%%%%%%%%%%%%%%%%%%%%%%%%%%%%%%%%%%%%
\section{Introduction} \label{sec:intro}

The improved infrastructure and an increase in the adoption of cyber-technology have led to increased connection and ease of interaction for users across the globe.  However, at the same time, these developments have increased users' exposure to risk. The importance of investing in security measures in this developing landscape is two-fold: while such expenditure helps entities protect their assets against security threats, by association it also benefits other interacting users, as an investing entity is less likely to be infected and used as a source of future attacks. In other words, a users' expenditure in security in an interconnected system provides \emph{positive externalities} to other users. Consequently, the provision of security is often studied as a problem of public good provision. In particular, when users are rational, the strategic decision making process leading to security investment decisions is studied as an \emph{(interdependent) security game} \cite{laszka12}. 

It is well-known that in an unregulated environment, the provision of public goods is in general inefficient \cite{mas-micro}.  To eliminate this inefficiency, the literature has proposed regulating mechanisms for implementing the socially optimal levels of security in these games, see e.g. \cite{laszka12, grossklags10, bohme10, walrand11}.  Specifically, examples of existing mechanisms in the literature include introducing subsidies and fines based on security investments \cite{kun03, grossklags10}, assessing rebates and penalties based on security outcomes \cite{grossklags10}, imposing a level of due care and establishing liability rules \cite{kun03, varian04}, etc. 

Our focus in the current paper is on mechanisms that use monetary payments/rewards to incentivize improved security behavior. Within this context, we will examine two incentive mechanisms, namely the \emph{Pivotal} \cite{clarke71} and \emph{Externality} \cite{hurwicz79} mechanisms, both of which induce socially optimal user behavior by levying a monetary tax on each user participating in the proposed mechanism. 

Aside from inducing optimal behavior, incentive mechanisms are often designed so as to maintain a \emph{(weakly) balanced budget (BB)} and ensure \emph{voluntary participation (VP)} by all users. The budget balance requirement states that the designer of the mechanism prefers to redistribute users' payments as rewards, and ideally to either retain a surplus as profit or at least to not sustain losses. Otherwise, the designer would need to spend external resources to achieve social optimality.  

The voluntary participation constraint on the other hand ensures that all users voluntarily take part in the proposed mechanism and the induced game, and prefer its outcome to that attained if they unilaterally decide to opt out of the mechanism. 
A user's decision when contemplating participation in an incentive mechanism is dependent not only on the structure of the game induced by the mechanism, but also on the options available when staying out. The latter is what sets the study of incentive mechanisms for security games apart from other public good problems where similar Pivotal and Externality mechanisms have been applied, e.g., \cite{parkes01, sharma11}. 

To elaborate on this underlying difference, we note that security is a \emph{non-excludable} public good. That is, although the mechanism optimizes the investments in a way that participating users are exposed to lower risks, those who stay out of the mechanism can benefit from the externalities of such improved state of security as well. 
The availability of these spill-overs in turn limits users' willingness to pay for the good or their interest in improving their actions. 
In contrast, with excludable public goods, e.g. transmission power allocated in a communication system \cite{sharma11}, users' willingness to participate is determined by the change in their utilities when contributing and receiving the good, as compared to receiving {\em no allocation at all}. This means that the designer has the ability to collect more taxes and require a higher level of contribution when providing an excludable good. 
As a result, tax-based mechanisms, such as the Externality mechanism (e.g. \cite{sharma11}) and the Pivotal mechanism (e.g. \cite{parkes01}), can be designed so as to incentivize the socially optimal provision of an excludable good, guarantee voluntary participation, and maintain (weak) budget balance. 

However, in this paper we show that given the non-excludable nature of security, there is no reliable tax-based mechanism that can achieve social optimality, voluntary participation, and (weak) budget balance simultaneously in all instances of security games. We show this result through two sets of counter-examples: we first limit the network structure to a star topology, and then consider the commonly studied weakest link model for users' risk functions. 
We then further elaborate on this particular nature of security games by examining the Pivotal and Externality mechanisms in the special case of a \emph{weighted total effort} interdependence model. This interdependence model is of particular interest as it can capture varying degrees and possible asymmetries in the influence of users' security decisions on one another. 
Specifically, we evaluate the effects of: (i) increasing users' self-dependence (equivalently, decreasing their interdependence), (ii) having two diverse classes of self-dependent and reliant users, and (iii) presence of a single dominant user, on the performance of the Pivotal and Externality mechanisms. We show that when possible, the selection of \rev{equilibria} that are less beneficial to the outliers helps the performance of both mechanisms, so that they can achieve optimality, budget balance, and voluntary participation simultaneously. In addition, we see that these incentive mechanisms become of interest when they can facilitate a tax-transfer scheme, such that users who are highly dependent on externalities pay to incentivize improved investments by others who are key to improving the state of security. 

{The main findings of this work can therefore be summarized as follows. First, we show that there is no tax-based incentive mechanism that can simultaneously guarantee social optimality, voluntary participation, and weak budget balance in all instances of security games. This result is applicable to other problems concerning the provision of non-excludable public goods over social and economic networks as well (see Section \ref{sec:related}). Second, we provide further insight on this impossibility by evaluating two incentive mechanisms, namely the Pivotal and Externality mechanisms, in weighted total effort games. We identify some of the parameters affecting the performance of these mechanisms, and instances in which the implementation of each mechanism is of interest.} 

The rest of this paper is organized as follows. We present the model for security games, as well as the Pivotal and Externality mechanisms, in Section \ref{sec:model}, followed by the general impossibility result in Section \ref{sec:imp}. Section \ref{sec:sim} illustrates this result by applying the Pivotal and Externality  mechanisms to weighted total effort models. We summarize related work in Section \ref{sec:related}, and conclude in Section \ref{sec:conclusion}.

\section{Security games: Model and Preliminaries} \label{sec:model}

\subsection{Model}
Consider a network of $N$ interdependent users. Each user $i$ can choose to exert effort towards securing its system, consequently achieving the \emph{level of security} or \emph{level of investment} $x_i\in \mathbb{R}_{\geq 0}$. Let $\mathbf{x}:=\{x_1, x_2, \ldots, x_N\}$ denote the \emph{state of security} of the system, i.e., the profile of security levels of all $N$ users. 

We let $h_i(\cdot):\mathbb{R}_{\geq 0}\rightarrow \mathbb{R}_{\geq 0}$ denote the \emph{investment cost function} of user $i$; it determines the monetary expenditure required to implement a level of security $x_i$. We assume this function is continuous, increasing, and convex. The assumption of convexity entails that security measures get increasingly costly as their effectiveness increases. 

The expected amount of assets user $i$ has subject to loss, given the state of security $\mathbf{x}$, is determined by the \emph{risk function}, and is denoted by $f_i(\cdot):\mathbb{R}^N_{\geq 0}\rightarrow \mathbb{R}_{\geq 0}$. We assume $f_i(\cdot)$ is continuous, non-increasing, and strictly convex, in all arguments $x_j$. The non-increasing nature of this function in arguments $x_j, j\neq i$, models the positive externality of users' security decisions on one another. The convexity on the other hand implies that the effectiveness of security measures in preventing attacks (or the marginal utility) is overall decreasing, as none of the available security measure can guarantee the prevention of all possible attacks. 

A user $i$'s \emph{(security) cost function} at a state of security $\mathbf{x}$ is therefore given by: 
\begin{align}
g_i(\mathbf{x}) = f_i(\mathbf{x}) + h_i(x_i)~.
\label{eq:gi}
\end{align}

We refer to the one-stage, full information game among the $N$ utility maximizing users with utility functions $u_i(\mathbf{x}) = -g_i(\mathbf{x})$ as the security game. The level of investments in the Nash equilibrium of these games, and their sub-optimality when compared to the socially optimal investments, has been extensively studied in the literature, see e.g. \cite{laszka12, walrand11, grossklags08}. Here, the socially optimal investment levels $\mathbf{x}^*$ are those maximizing the total welfare, or equivalently, minimizing the sum of all users' costs, i.e.,  
\begin{eqnarray}
\mathbf{x}^* = \arg\min_{\mathbf{x}\succeq 0} \sum_i g_i(\mathbf{x}). 
\label{eq:opt}
\end{eqnarray} 
The literature has further proposed mechanisms for decreasing the inefficiency gap in security games, by either incentivizing or dictating improved security investments; see \cite{laszka12} for a survey. 
Our focus in the present paper is on regulating mechanisms that use monetary taxation to incentivize socially optimal security behavior. Such mechanisms assess a tax $t_i$ on each user $i$; this tax may be positive, negative, or zero, indicating payments, rewards, or no transaction, respectively. 

We further assume that users' utilities are quasi-linear. Therefore, the \emph{total cost} of a user $i$ when it is assigned a tax $t_i$ is given by:   
\begin{eqnarray}
\mathrm{g}_i(\mathbf{x}, t_i) := g_i(\mathbf{x})+t_i. 
\label{eq:tot_g}
\end{eqnarray}

In addition to implementing the socially optimal solution, incentive mechanisms are often required to satisfy two desirable properties. First, when using taxation, the mechanism designer prefers to maintain \emph{(weak) budget balance (BB)}; i.e., it is desirable to have %\begin{eqnarray*}
$\sum_i t_i \geq 0$. 
%\end{eqnarray*} 
In particular, $\sum_i t_i<0$ implies a budget deficit, such that the implementation of the mechanism would call for the injection of additional resources by the designer.

In addition, it is desirable to design the mechanism in a way that users' \emph{voluntary participation (VP)} conditions are satisfied; i.e. users prefer implementing the socially optimal outcome while being assigned taxes $t_i$, to the outcome attained had they unilaterally opted out. Otherwise, the designer would need to enforce initial cooperation in the mechanism. Note that we focus on the notion of voluntary participation instead of the usual \emph{individual rationality (IR)} constraint, \rev{which requires a user to prefer participation to the outcome it attained at the state of anarchy (i.e., prior to the implementation of the mechanism)}. As mentioned in Section \ref{sec:intro}, such distinction is important as security is a non-excludable public good, i.e., users can still benefit from the externalities of the actions of users participating in the mechanism, even when opting out themselves. This is in contrast to games with excludable public goods, where VP and IR are equivalent. 

We now proceed to introduce the Pivotal and Externality tax-based incentive mechanisms for security games.

%%%%%%%%%%%%%%%%%%%%%%%%%%%%%%%%%%%%%%%%%%%
\subsection{The Pivotal Mechanism} \label{sec:pivotal}
Groves mechanisms \cite{mas-micro, parkes01}, also commonly known as Vickery-Clarke-Groves (VCG) mechanisms, refer to a family of mechanisms in which, through the appropriate design of taxes for users with quasi-linear utilities, a mechanism designer can incentivize users to reveal their true preferences in dominant strategies, thus implementing the socially optimal solution. However, the (weak) budget balance and voluntary participation conditions do not necessarily hold in these mechanisms, and are further dependent on the specifics of the design, as well as the game environment. 

In general, let $u_i(k, \theta_i, t_i) = v_i(k, \theta_i) - t_i$ be user $i$'s utility. Here, $\theta_i$ is user $i$'s type; a user's type determines the preference of the user over possible outcomes. In security games, a user $i$'s type is its risk and investment cost functions $\{f_i(\cdot), h_i(\cdot)\}$, or equivalently, its cost function $g_i(\cdot)$. Users are required to report their types to the mechanism designer, based on which the designer decides on an allocation $k$. In security games, an allocation is the vector of investments $\mathbf{x}$ prescribed by the mechanism. 

The VCG family of mechanisms achieve truth revelation and efficiency by assigning the following taxes to users, when their reported types are $\hat{\theta}$: 
\[t_i(\hat{\theta}) =  \alpha_i(\hat{\theta}_{-i}) - \sum_{j\neq i} v_j(k^*(\hat{\theta}), \hat{\theta}_j)~. \]
Here, $k^*(\hat{\theta}) = \arg\max_k \sum_i v_i(k, \hat{\theta}_i)$ is the socially optimal allocation given users' reported types, and $\alpha_i(\cdot)$ is an arbitrary function that depends on the reported types of users other than $i$. Any choice of this function results in truth revelation and a socially efficient outcome, and a careful design may further result in VP and/or (W)BB. 

One such choice that can achieve VP in certain environments is the \emph{Pivotal}, or \emph{Clarke}, mechanism \cite{clarke71, parkes01}, with taxes given by: 
\[t_i(\hat{\theta}) =  \sum_{j\neq i} v_j(k_{-i}^*(\hat{\theta}_{-i}), \hat{\theta}_j) - \sum_{j\neq i} v_j(k^*(\hat{\theta}), \hat{\theta}_j)~. \]
Here, $k_{-i}^*(\hat{\theta}_{-i}) = \arg\max_k \sum_{j\neq i} v_j(k, \hat{\theta}_j)$, is the outcome maximizing the social welfare in the absence of user $i$. This mechanism satisfies the participation constraints and achieves weak budget balance in many private and public good games \cite{parkes01}; however, this is not necessarily the case in security games. 

The taxes in the Pivotal mechanism for the security game can be set as follows: 
\begin{align}
t^P_i = \sum_{j\neq i} g_j(\mathbf{x}^*_{-i}, {x}^*_i) - \sum_{j\neq i} g_j(\hat{\mathbf{x}}^i_{-i}, \hat{x}^i_i)~,
\label{eq:vcg-tax}
\end{align}
where, $g_i(\mathbf{x})$ is user $i$'s security cost function, \rev{$\mathbf{x}^* := (\mathbf{x}^*_{-i}, {x}^*_i)  = \arg\min_{\mathbf{x}} \sum_i g_i(\mathbf{x})$} is the socially optimal solution, and $\hat{\mathbf{x}}^i_{-i}$ is the cost minimizing actions of users $j\neq i$ given user $i$'s action $\hat{x}^i_i$, and is determined by $\hat{\mathbf{x}}^i_{-i} = \arg\min_{\mathbf{x}_{-i}} \sum_{j\neq i} g_j(\mathbf{x}_{-i}, \hat{x}^i_i)$. In a game of complete information, $\mathbf{\hat{x}}^i$ will be the Nash equilibrium of the game between user $i$ and the $N-1$ participating users. 

It is straightforward to verify that this design of the Pivotal mechanism in security games internalizes the externalities of users' actions, and can thus lead to the implementation of the socially optimal solution. Formally, 
\begin{proposition}\label{prop1}
In the Pivotal mechanism with taxes given by \eqref{eq:vcg-tax}, investing the socially optimal level of investment ${x}^*_i$ will be individually optimal, for all users $i$. Therefore, the socially optimal solution is implemented. 
\end{proposition} 
Furthermore, such design will ensure participation by all users. That is, 
\begin{proposition}\label{prop2}
The Pivotal mechanism with taxes given by \eqref{eq:vcg-tax} satisfies all voluntary participation constraints. 
\end{proposition}

The proofs of these propositions follow directly from existing literature, see e.g. \cite{parkes01}.

%%%%%%%%%%
\subsection{The Externality Mechanism} \label{sec:externality}

We next examine a taxation mechanism that can achieve the socially optimal solution in security games, while maintaing a balanced budget. This mechanism is adapted from \cite{hurwicz79}.   
The components of the mechanism are as follows. 

{\em The message space:} 
Each user $i$ provides a message ${m}_i:=(\boldsymbol{\chi}_i, \boldsymbol{\pi}_i)$ to the mechanism designer. $\boldsymbol{\chi}_i \in \mathbb{R}^N$ denotes user $i$'s proposal on the public good, i.e., it proposes the amount of security investment to be made by everyone in the system, referred to as an {\em investment profile}. 

$\boldsymbol{\pi}_i\in \mathbb{R}^N_+$ denotes a {\em pricing profile} which suggests the amount to be paid by everyone.  As illustrated below, this is used by the designer to determine the taxes of all users. Therefore, the pricing profile is user $i$'s proposal on the private good.

{\em The outcome function:}
The outcome function takes the message profiles $\mathbf{m}:=\{m_1, m_2, \ldots, m_N\}$ as input, and determines the security investment profile $\mathbf{\hat{x}}$ and a {tax} profile $\mathbf{\hat{t}}^E$ as follows:
\begin{align}
\mathbf{\hat{x}}(\mathbf{m}) &= \frac{1}{N} \sum_{i=1}^N \boldsymbol{\chi}_i~, ~~~~ \label{eq:out_x} \\
{\hat{t}}^E_i(\mathbf{m}) &= (\boldsymbol{\pi}_{i+1} - \boldsymbol{\pi}_{i+2})^T\mathbf{\hat{x}}(\mathbf{m}) + (\boldsymbol{\chi}_i - \boldsymbol{\chi}_{i+1})^T \text{diag}(\boldsymbol{\pi}_i)(\boldsymbol{\chi}_{i} - \boldsymbol{\chi}_{i+1}) ~~~~ \notag\\
 & -  (\boldsymbol{\chi}_{i+1} - \boldsymbol{\chi}_{i+2})^T \text{diag}(\boldsymbol{\pi}_{i+1})(\boldsymbol{\chi}_{i+1} - \boldsymbol{\chi}_{i+2}), \forall i.
\label{eq:out_t}
\end{align}
In \eqref{eq:out_t}, $N+1$ and $N+2$ are treated as $1$ and $2$, respectively.  

Note that as $\sum_i \hat{t}^E_i = 0$ by \eqref{eq:out_t}, the budget balance condition is satisfied through this construction.  What this means is that the designer will not be spending resources or making profit, as the users whose tax $\hat{t}_i$ is positive will be financing the rewards for those who have negative taxes. In other words, the mechanism proposes a tax \emph{redistribution} scheme to incentivize improved security investments.  

To establish that the Externality mechanism can implement the socially optimal outcome in security games, we first need to show that a profile $(\mathbf{\hat{x}}(\mathbf{{m}^*}),\mathbf{\hat{t}}^E(\mathbf{{m}^*}))$, derived at any possible NE $\mathbf{{m}^*}$ of the Externality regulated game, is the socially optimal solution. Formally, 
\begin{theorem} \label{th1}
Let  $(\mathbf{\hat{x}}(\mathbf{{m}^*}),\mathbf{\hat{t}}^E(\mathbf{{m}^*}))$ be the investment and tax profiles obtained at the Nash equilibrium $\mathbf{{m}^*}$ of the regulated security game. Then, $\mathbf{\hat{x}}$ is the optimal solution to the centralized problem \eqref{eq:opt}. Furthermore, if $\mathbf{\bar{m}}$ is any other Nash equilibrium of the proposed game, then $\mathbf{\hat{x}}(\mathbf{\bar{m}}) = \mathbf{\hat{x}}(\mathbf{{m}^*})$.
\end{theorem}
Furthermore, we have to show the converse of the previous statement, i.e., given an optimal investment profile, there exists an NE of Externality regulated game which implements this solution. Formally, we can show the following: 
\begin{theorem} \label{th2}
Let $\mathbf{x}^*$ be the optimal investment profile in the solution to the centralized problem \eqref{eq:opt}. Then, there exists at least one Nash equilibrium $\mathbf{{m}^*}$ of the regulated security game such that $\mathbf{\hat{x}}(\mathbf{{m}^*})=\mathbf{x}^*$.
\end{theorem}

The proofs of these theorems follow the method used by \cite{hurwicz79, sharma11}.  We refer the interested reader to these papers, as well as our earlier work \cite{naghizadeh14b}, where we present a sketch of the proof of Theorem \ref{th1}, along with an intuitive interpretation for this mechanism. Using the proof of \ref{th1}, we show that the tax terms $t_i^E$ at the equilibrium of the Externality mechanism are given by: 
\begin{align}
t_i^E(\mathbf{x}^*) = -\sum_j x_j^* \frac{\partial f_i}{\partial x_j}(x^*) - \frac{\partial h_i}{\partial x_i}x_i^*~.
\label{eq:ext-tax}
\end{align}
The interpretation is that by implementing this mechanism, each user $i$ will be financing part of user $j\neq i$'s reimbursement. According to \eqref{eq:ext-tax}, this amount is proportional to the positive externality of $j$'s investment on user $i$'s utility.

\section{An impossibility result} \label{sec:imp}

In the previous section, we stated two well-known tax-based incentive mechanisms for incentivizing socially optimal actions, namely the Pivotal and the Externality mechanisms, in the context of security games. The Pivotal mechanism is designed to guarantee voluntary participation, while the Externality mechanism focuses on budget balance. Following these observations, one may ask whether either of these schemes, or other tax-based mechanisms, can achieve social optimality, while guaranteeing both budget balance and voluntary participation simultaneously, in all instances of security games. In this section, we show that in fact no such reliable mechanism exists. We illustrate this impossibility through two families of counter-examples. The first counter-example considers games in which the network structure is a star topology, while the second family focuses on security games with weakest link risk functions. 

In what follows, to evaluate users' voluntary participation conditions, we consider a user $i$, referred to as the \emph{loner} or \emph{outlier}, who is unilaterally contemplating opting out of this mechanism. As the game considered here is one of full information, the remaining participating users, who are choosing a welfare maximizing solution for their ($N-1$)-user system, will have the ability to predict the best-response of the loner to their collective action, and thus choose their investments accordingly. As a result, the equilibrium investment profile when user $i$ opts out is the Nash equilibrium of the game between the $N-1$ participating users and this loner. We will henceforth refer to this equilibrium as the \emph{exit equilibrium} (EE).

%%%%%%%%%%%%%%%%%%%%%%%%%%%%%%%%%%%%%%%%%%%%
\subsection{Counter-example I: the star topology}
\begin{figure}
\centering
\begin{tikzpicture}[level distance=1.5cm, grow=down,
    every node/.style={draw, circle, thin},
    norm/.style ={edge from parent/.style={thick, draw}},
    ndots/.style ={edge from parent/.style={}}
]
\node (1) {1}
    child[norm] {node (2) {2}}
        child[norm] {node (3) {3}}
        child[ndots] {node (A) [draw=none] {$\ldots$}}
        child[norm] {node (N) {N}}
    ;
\end{tikzpicture}
\caption{No tax-based mechanism can guarantee SO, BB, and VP, in a star topology}
\label{ex:star}
\end{figure}
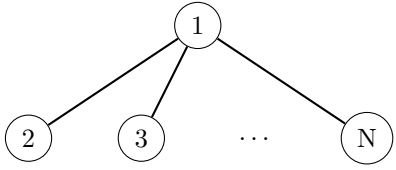

Assume some tax-based incentive mechanism $\mathcal{M}$ is proposed for security games. 
Consider $N$ users connected through the star topology depicted in Fig. \ref{ex:star}, where the security decisions of the root affects all leaves, but each leaf's investment only affects itself and the root. Formally, let the cost function of the root be given by:
\[g_1(\mathbf{x}) = f(x_1+\sum_{j=2}^N x_j) + cx_1~,\]
and that of all leaves $j\in\{2,\ldots, N\}$ be:
\[g_j(\mathbf{x}) = f(x_1+x_j) + cx_j~.\]
Here, $f(\cdot)$ is any function satisfying the assumptions in Section \ref{sec:model}. The investment cost functions $h_i(\cdot)$ are linear, with the same unit investment cost $c$ for all users. 
To find the socially optimal investment profile, we solve the optimization problem of minimizing the sum of all users' costs, $G(\mathbf{x}):=\sum_{i=1}^N g_i(\mathbf{x})$, subject to non-negative user investments. This profile, $\mathbf{x}^*$, should satisfy: 
\begin{align*}
&\frac{\partial f}{\partial x}({x}^*_1+\sum_{j=2}^N x^*_j) + \sum_{j=2}^N \frac{\partial f}{\partial x}({x}^*_1+{x}^*_j) +c - \lambda_1 = 0, ~~ \lambda_1x_1^* = 0~,\\
&\frac{\partial f}{\partial x}({x}^*_1+\sum_{j=2}^N x^*_j) + \frac{\partial f}{\partial x}({x}^*_1+ {x}^*_j) +c -\lambda_j = 0,~~ \lambda_jx_j^* = 0, ~~ \forall j=2, \ldots, N~. 
\end{align*}
Based on the above, it is easy to see that in the socially optimal investment profile for this graph, only the root will be investing in security, while all leaves free-ride on the resulting externality. This socially optimal investment profile $\mathbf{x}^*$ is given by: 
\[\frac{\partial f}{\partial x}(x_1^*) = -{\frac{c}{N}}, \quad x_j^*=0, \forall j=2, \ldots, N~.\]

Now, assume the root user is considering stepping out of the mechanism. To find the investment profile $\mathbf{\hat{x}}^1$ resulting from this unilateral deviation, first note that the leaves' security decisions will not affect one another, so that the socially optimal investment profile for the $N-1$ leaves is the same as their individually optimal decisions. User 1 will also be choosing its individually optimal level of investment. Therefore, using users' first order conditions for cost minimization, the exit equilibrium $\mathbf{\hat{x}}^1$ 
%will satisfy: 
%\begin{align*}
%&\frac{\partial f}{\partial x}(\hat{x}^1_1)+c \geq 0\\
%&\frac{\partial f}{\partial x}(\hat{x}^1_1+\hat{x}^1_j)+c \geq 0,~~ \forall j=2, \ldots, N~.
%\end{align*}
%We conclude that the exit equilibrium when user 1 unilaterally leaves the Pivotal mechanism 
is: 
\[\frac{\partial f}{\partial x}(\hat{x}_1^1) = -{c}, \quad \hat{x}^1_j=0, \forall j=2, \ldots, N~.\]

Finally, if any leaf user $j\in\{2, \ldots, N\}$ leaves the mechanism, the exit equilibrium $\mathbf{\hat{x}}^j$ will satisfy: 
\begin{align*}
&\frac{\partial f}{\partial x}(\hat{x}^j_1+\sum_{j=2}^N \hat{x}^j_i) + \sum_{k\neq 1,j} \frac{\partial f}{\partial x}(\hat{x}^j_1+{x}^j_k) +c - \lambda_1 = 0, ~~ \lambda_1\hat{x}_1^j=0~,\\
&\frac{\partial f}{\partial x}(\hat{x}^j_1+\sum_{j=2}^N \hat{x}^j_i) + \frac{\partial f}{\partial x}(\hat{x}^j_1+ \hat{x}^j_k) +c - \lambda_k = 0, ~~ \lambda_k\hat{x}_k^j=0, ~~ \forall k=2, \ldots, N,~ k\neq j\\ 
&\frac{\partial f}{\partial x}(\hat{x}^j_1+ \hat{x}^j_j)+c - \lambda_j = 0, ~~ \lambda_j\hat{x}_j^j=0,~.
\end{align*}
Again, it is easy to see that $\hat{x}^j_k=0, \forall k\in\{2, \ldots, N\}$. Therefore, the exit equilibrium when user $j\in\{2, \ldots, N\}$ unilaterally leaves the mechanism is given by: 
\[\frac{\partial f}{\partial x}(\hat{x}_1^j) = -\frac{c}{N-1}, \quad \hat{x}^j_k=0, \forall k=2, \ldots, N~.\]

We now use the socially optimal investment profile and the exit equilibria to evaluate voluntary participation and budget balance in a general mechanism $\mathcal{M}$. Assume $\mathcal{M}$ assigns a tax $t_i^*$ to a participating user $i$. Then, voluntary participation will hold if and only if \rev{$\mathrm{g}_i(\mathbf{x}^*, t_i^*) \leq g_i(\mathbf{\hat{x}}^i), ~\forall i$}, which reduces to: 
\begin{align*}
t_1^* &\leq f(\hat{x}^1_1) - f(x_1^*) + c(\hat{x}^1_1-x_1^*)~, \notag\\
t_j^* &\leq f(\hat{x}^j_1) - f(x_1^*), ~\forall j\in\{2, \ldots, N\}~.
\label{eq:vp_star}
\end{align*} 
The sum of these taxes is thus bounded by: 
\begin{align*}
\sum_{i=1}^N t^*_i &\leq f(\hat{x}^1_1) - f(x_1^*) + c(\hat{x}^1_1-x_1^*) + (N-1)(f(\hat{x}^j_1) - f(x_1^*))
%\\
%& = N\left(f(\hat{x}^j_1) - f(x_1^*)\right) + c(\hat{x}^j_1-x_1^*) + f(\hat{x}_1^1) - f(\hat{x}^j_1) + c(\hat{x}^1_1-\hat{x}^j_1)~. 
\end{align*}
However, the above sum could be negative, e.g., when $f(z)=\exp(-z)$ or $f(z)=\frac{1}{z}$, indicating that weak budget balance will fail regardless of how the taxes are determined in a mechanism $\mathcal{M}$.

%%%%%%%%%%%%%%%%%%%%%%%%%%%%%%%%%%%%%%%%%%%%
\subsection{Counter-example II: weakest-link games}
In this section, we again assume a general tax-based incentive mechanism $\mathcal{M}$ is proposed for the security games. We focus on a family of security games which approximate the \emph{weakest link} risk function $f_i(\mathbf{x}) = \exp(-\min_j{x_j})$ \cite{varian04, laszka12}. Intuitively, this model states that an attacker can compromise the security of an interconnected system by taking over the least protected node. To use this model in our current framework, we need a continuous, differentiable approximation of the minimum function. We use the approximation $\min_j{x_j}\approx -\frac{1}{\rho}\log\sum_{j}\exp(-\rho x_j)$, where the accuracy of the approximation is increasing in the constant $\rho>0$. User $i$'s cost function is thus given by:
\[{g}_i(\mathbf{x}) = (\sum_{j=1}^N \exp(-\rho x_j))^{1/\rho} + cx_i~,\]
where investment cost functions $h_i(\cdot)$ are assumed to be linear, with the same unit investment cost $c$ for all users. 

In this game, the socially optimal investment profile $\mathbf{x}^*$ is given by the solution to the first order condition $\frac{\partial \sum_j g_j(\mathbf{x})}{\partial x_i} = 0$, which leads to: 
\[N \exp(-\rho x_i^*) (\sum_{j=1}^N \exp(-\rho x^*_j))^{\frac{1}{\rho} - 1} = c~, \forall i.\] 
By symmetry, all users will be exerting the same socially optimal level of effort: 
\[x_i^* = \frac{1}{\rho} \ln\frac{N}{c^\rho}~, \forall i~.\]

Next, assume a user $i$ unilaterally steps out of the mechanism, while the remaining users continue participating. The exit equilibrium profile $\mathbf{\hat{x}}^i$ can be determined using: 
\begin{align*}
(N-1) \exp(-\rho \hat{x}^i_j) (\sum_{k\neq i} \exp(-\rho \hat{x}^i_k)+ \exp(-\rho \hat{x}^i_i))^{\frac{1}{\rho} - 1} = c~,\\
\exp(-\rho \hat{x}^i_i) (\sum_{k\neq i} \exp(-\rho \hat{x}^i_k)+ \exp(-\rho \hat{x}^i_i))^{\frac{1}{\rho} - 1} = c~. 
\end{align*}
Solving the above, we get: 
\begin{align*}
\hat{x}^i_i &= \frac{1}{\rho} \ln\frac{2^{1-\rho}}{c^\rho}~\\
\hat{x}^i_j &= \frac{1}{\rho} \ln\frac{(N-1)2^{1-\rho}}{c^\rho}~, \forall j\neq i~. 
\end{align*}

We now use the socially optimal investment profile and the exit equilibria to analyze users' participation incentives in a general mechanism $\mathcal{M}$, as well as the budget balance conditions. Denote by $t_i^*$ the tax assigned to user $i$ by $\mathcal{M}$. A user $i$'s total cost functions when participating and staying out are given by: 
\begin{align*}
\mathrm{g}_i(\mathbf{x}^*, t_i^*)&= (N\exp(-\rho x_i^*))^{1/\rho} + cx_i^* + t_i^* = c(1+x_i^*) + t_i^*\\
g_i(\mathbf{\hat{x}}^i) &= (\exp(-\rho \hat{x}^i_i)+(N-1)\exp(-\rho \hat{x}^i_j))^{1/\rho} + c\hat{x}^i_i = c(2+\hat{x}^i_i)~. 
\end{align*}
The voluntary participation condition for this user will hold if and only if $\mathrm{g}_i(\mathbf{x}^*, t_i^*) \leq g_i(\mathbf{\hat{x}}^i)$, which reduces to: 
\begin{align}
c(1+x_i^*) + t_i^* \leq c(2+\hat{x}^i_i) \Leftrightarrow t_i^* \leq c(1+\frac{1}{\rho}\ln\frac{2^{1-\rho}}{N})~.
\label{eq:vp_weakest}
\end{align} 
On the other hand, for weak budget balance to hold, we need $\sum_i t_i^* \geq 0$. Nevertheless, by \eqref{eq:vp_weakest}, we have: 
\[\sum_i t_i^* \leq cN(1+\frac{1}{\rho}\ln\frac{2^{1-\rho}}{N})~.\]
It is easy to see that given $\rho$ and for any $N>e^{\rho} 2^{1-\rho}$, the above sum will always be negative, indicating a budget deficit for a general mechanism $\mathcal{M}$, regardless of how taxes are determined. \footnote{It is worth mentioning that when $N\leq e^{\rho} 2^{1-\rho}$, the Externality mechanism with taxes determined by \eqref{eq:ext-tax} will be budget balanced and guarantee voluntary participation. However, the Pivotal mechanism will carry a deficit in both regions.}

%%%%%%%%%%%%%%%%%%%%%%%%%%%%%%%%%%%%%%%%%%%%
\subsection{A note on the nature of this impossibility result}
To close this section, we would like to point out that the impossibility result on a simultaneous guarantee of social optimality, voluntary participation, and weak budget balance, is demonstrated through two family of counter-examples. In other words, we have shown that without prior knowledge of the graph structure or users' preferences, it is not possible for a designer to propose a \emph{reliable} mechanism; that is, one which can promise to achieve SO, VP, and WBB, regardless of the realizations of utilities.  Nevertheless, it may still be possible to design reliable mechanisms under a restricted space of problem parameters; in fact we identify a few such instances in Section \ref{sec:sim} by analyzing the class of weighted total effort models.

\section{Weighted total effort models: analysis and simulation} \label{sec:sim}

In the remainder of the paper, to further illustrate some of the parameters affecting the performance of incentive mechanisms in security games, we focus on the Pivotal and Externality mechanisms. We consider the special case of weighted total effort games, and identify some of the factors that affect the total budget and participation incentives in the Pivotal and Externality mechanisms, respectively. 

%%%%
\subsection{Choice of the risk function} \label{sec:lit-wte}

The gap between the Nash equilibrium and the socially optimal investment profile of a security game, as well as users' participation incentives and possible budget imbalances, are dependent on the specifics of the security cost functions defined in \eqref{eq:gi}. In particular, an appropriate choice of the risk functions $f_i(\cdot)$ for a given game is based on factors such as the type of interconnection, the extent of interaction among users, and the type of attack. Several models of security interdependency have been proposed and studied in the literature; these include the \emph{total effort}, \emph{weakest link}, and \emph{best shot} models considered in the seminal work of Varian on security games \cite{varian04}, as well as the \emph{weakest target} games proposed in \cite{grossklags08}, the \emph{effective investment} and \emph{bad traffic} models in \cite{walrand11}, and the \emph{linear influence network} games in \cite{miura08}. 

In this paper, we take the special case of the \emph{weighted total effort} games, with exponential risks and linear investment cost functions, to study the effects of interdependency on the performance of the Pivotal and Externality mechanisms. Formally, the total cost function of a user $i$ in this model is given by: 
\begin{align}
\mathrm{g}_i(\mathbf{x}, t_i) = \exp(-\sum_{j=1}^N a_{ij}{x_j}) + cx_i + t_i~.
\label{eq:u_wte}
\end{align}
Here, the investment cost function is assumed linear, $h_i(x_i)=cx_i$. The coefficients $a_{ij}\geq 0$ determine the dependence of user $i$'s risk on user $j$'s action. Consequently, user $i$'s risk is dependent on a weighted sum of all users' actions. 

In particular, to isolate the effect of different features of the model on the performance of the two mechanisms, we focus on three sub-classes of the weighted total effort model. We first look at the effects of varying users' self-dependence. Next, we consider the effects of diversity, by breaking users into two groups of self-dependent and reliant users. Finally, we study the effect of making all users increasingly dependent on a single node's action. We present numerical results and intuitive interpretation for each of the above scenarios; formal analysis is given in the online appendix.

%%%%%%%%%%%%%%%%%%%%%%
%%%%%%%%%%%%%%%%%%%%%%
%%%%%%%%%%%%%%%%%%%%%%
%%%%%%%%%%%%%%%%%%%%%%
\subsection{Effects of self-dependence} \label{sec:sim_adiag}
Consider a collection of $N$ users, with total cost functions determined according to \eqref{eq:u_wte}, with $a_{ii} = a, \forall i$, and $a_{ij}=1, \forall i,j\neq i$: %The total cost functions of the users will then be given by: 
\[\mathrm{g}_i(\mathbf{x}, t_i) = \exp(-a x_i - \sum_{j\neq i} x_j) + c x_i + t_i~.\]
We assume $c<a$, so as to ensure the existence of non-zero equilibria; i.e., at least one user exerts non-zero effort at any equilibrium of the game. The socially optimal and exit equilibria of this game can be determined by using the first order conditions on the users' cost minimization problems, subject to non-negative investments. The resulting systems of equations can be solved to determine the possible exit equilibria, as well as parameter conditions under which each equilibrium happens; the results are summarized in Table \ref{t:vpbb_adiag}.   

According to this table, we can identify five sets of parameter conditions under which different exit equilibria are possible. We can further analyze each case separately to find whether the voluntary participation conditions are satisfied under the Externality mechanism, as well as whether the Pivotal mechanism can operate without a budget deficit. 
%The full analysis is presented in {Appendix B}, 
These results are summarized in Table \ref{t:vpbb_adiag} as well. 
\begin{table}
\centering
\caption{Can VP and BB hold simultaneously? -  effect of self-dependence}
\begin{center}
\begin{tabular}{|c|p{2.2cm}|p{3.6cm}|p{1.6cm}|p{1.2cm}|}
\cline{2-5}
\multicolumn{1}{c|}{} & Exit \newline Equilibrium & Parameter Conditions & VP in \newline Externality & BB in Pivotal\\
\hline
\textbf{CASE $\alpha$} & $\hat{x}^i_i=0, \hat{x}^i_j>0$ & $a>1$, and \newline $(1+\frac{N-2}{a})^{N-1}>(\frac{a}{c})^{a-1}$ & Never & Never \\
\hline
\textbf{CASE $\beta$} & $\hat{x}^i_i>0, \hat{x}^i_j>0$ &$a>1$, and \newline $(1+\frac{N-2}{a})^{N-1}<(\frac{a}{c})^{a-1}$ & Never & Never \\
\hline
\textbf{CASE $\gamma$} & $\hat{x}^i_i=0, \hat{x}^i_j>0$ &$a<1$, and $\forall N,c$ & Never & Never \\
\hline
\textbf{CASE $\omega$} & $\hat{x}^i_i>0, \hat{x}^i_j=0$ &$a<1$, and  \newline $(1+\frac{N-2}{a})^a < (\frac{a}{c})^{1-a}$& Always & Always \\
\hline
\textbf{CASE $\zeta$} & $\hat{x}^i_i>0, \hat{x}^i_j>0$ &$a<1$, and \newline $(1+\frac{N-2}{a})^a < (\frac{a}{c})^{1-a}$& Always & Always \\
\hline
\end{tabular}
\end{center}
\label{t:vpbb_adiag}
\end{table}%

\subsubsection{Simulations: cases $a>1$}
As seen in Table \ref{t:vpbb_adiag}, when $a>1$, neither of the two mechanisms can maintain a balanced budget and guarantee voluntary participation simultaneously, in either of the realized equilibria. In this section, we further examine the effect of changing $a$, $N$, and $c$ on the mechanisms' performance. In particular, we plot the sum of all taxes, $\sum_i t_i^P$, in the Pivotal mechanism. For the Externality mechanism, we plot $g_i(\mathbf{\hat{x}}^i) - \mathrm{g}_i(\mathbf{x^*}, t_i^E)$ per user $i$; i.e., the benefit of participation (in terms of cost reduction) for that user. We also consider the effect of these changes on the price of anarchy of the security game, by looking at the ratio of sum of the costs at the symmetric Nash equilibrium, over the sum of the costs at the socially optimal solution. 

\begin{figure}
\centering
\begin{minipage}[t]{0.49\columnwidth}
\includegraphics[width=\textwidth]{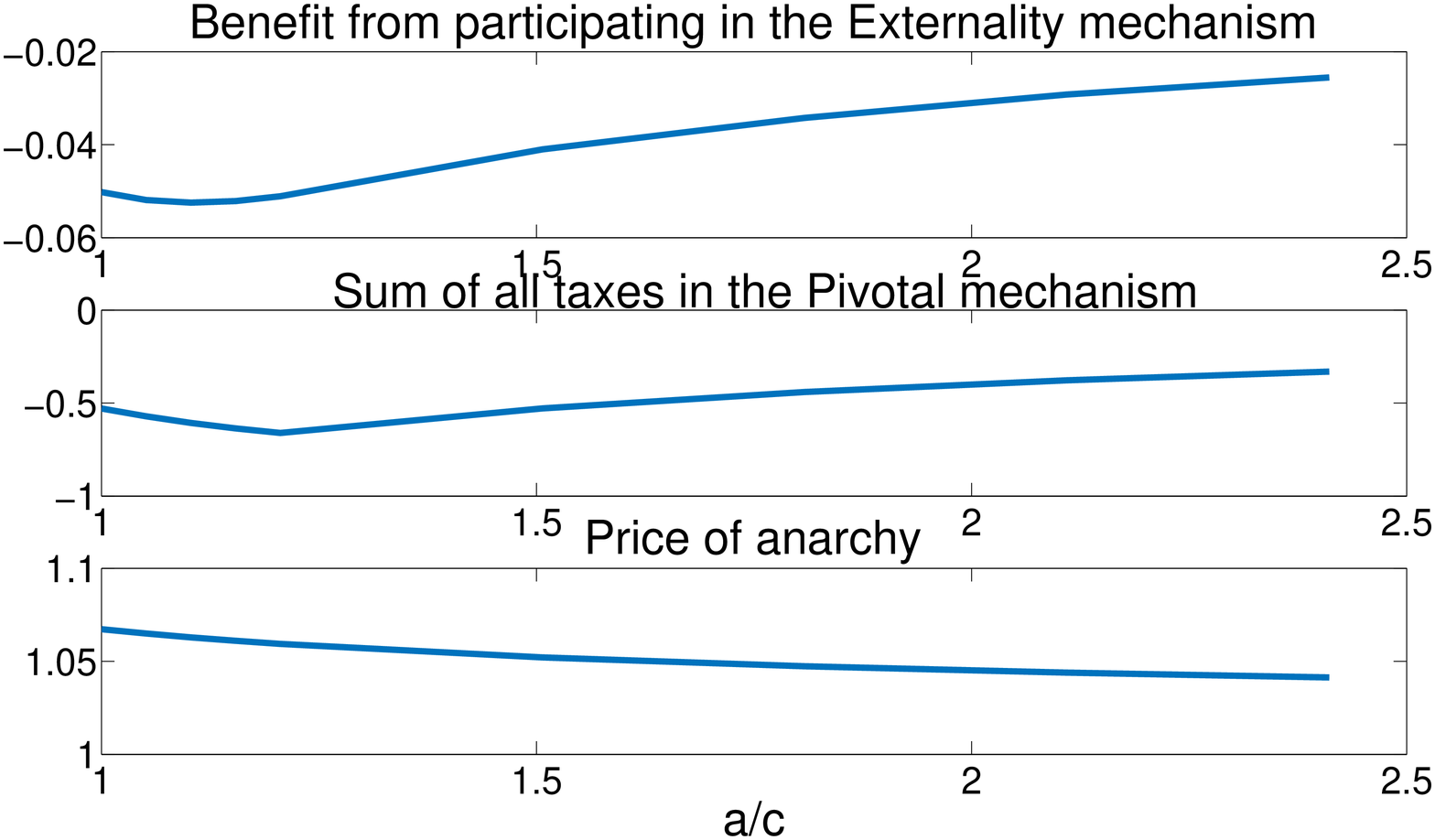}
\caption{Decreasing investment cost $c$}
\label{plot_1_c}
\end{minipage}
\begin{minipage}[t]{0.49\columnwidth}
\includegraphics[width=\textwidth]{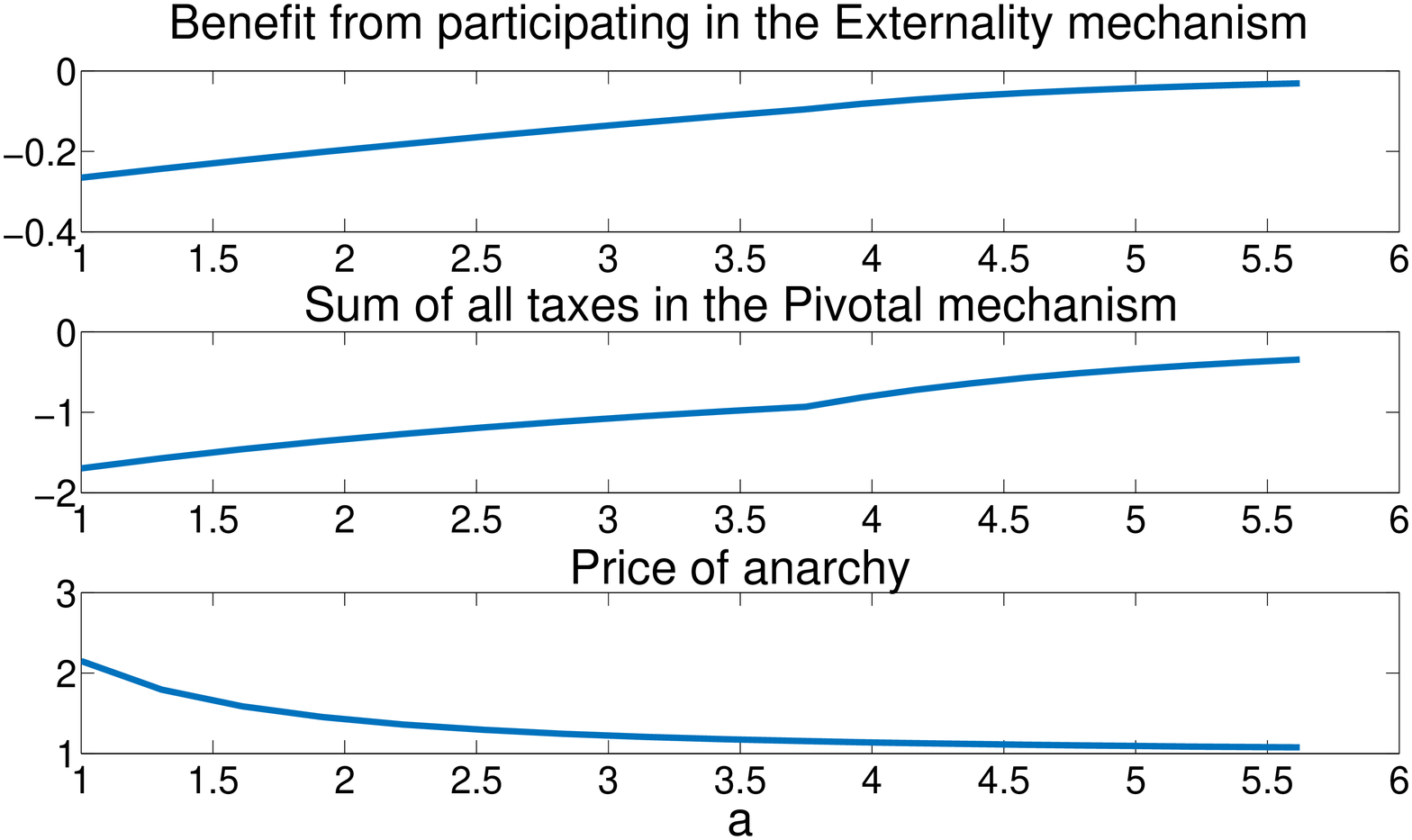}
\caption{Increasing self-dependence $a$}
\label{plot_2_a}
\end{minipage}
\end{figure}

\paragraph{Changing $c$}
In order to understand the effect of the unit cost, we set $N=6$ and $a=10$. We then change the fraction $a/c$, so that initially we are in [Case $\alpha$]: $\hat{x}^i_i=0, \hat{x}^i_j>0$, and gradually move to [Case $\beta$]: $\hat{x}^i_i>0, \hat{x}^i_j>0$. Intuitively, we will be gradually reducing the unit cost of investment, so that the outlier finds it efficient to continue investing even when leaving the mechanism. Figure \ref{plot_1_c} illustrates the results. 

%\begin{figure}
%\centering
%\includegraphics[width=0.7\textwidth]{Plots/Plot_1}
%\caption{Effect of decreasing the investment cost $c$}
%\label{plot_1_c}
%\end{figure} 

\paragraph{Changing $a$}
We next set $N=6$ and $c=1$. We then increase $a$, starting from $a=1$, so that initially we are in [Case $\alpha$]: $\hat{x}^i_i=0, \hat{x}^i_j>0$, and gradually move to [Case $\beta$]: $\hat{x}^i_i>0, \hat{x}^i_j>0$. Intuitively, we are gradually increasing self-dependence, and therefore making a unit of investment more effective for the user, so that outliers exert non-zero effort. Figure \ref{plot_2_a} illustrates the results. 

%\begin{figure}
%\centering
%\includegraphics[width=0.7\textwidth]{Plots/Plot_2}
%\caption{Increasing self-dependence $a$}
%\label{plot_2_a}
%\end{figure} 

\begin{figure}[b]
\centering
\includegraphics[width=0.7\textwidth]{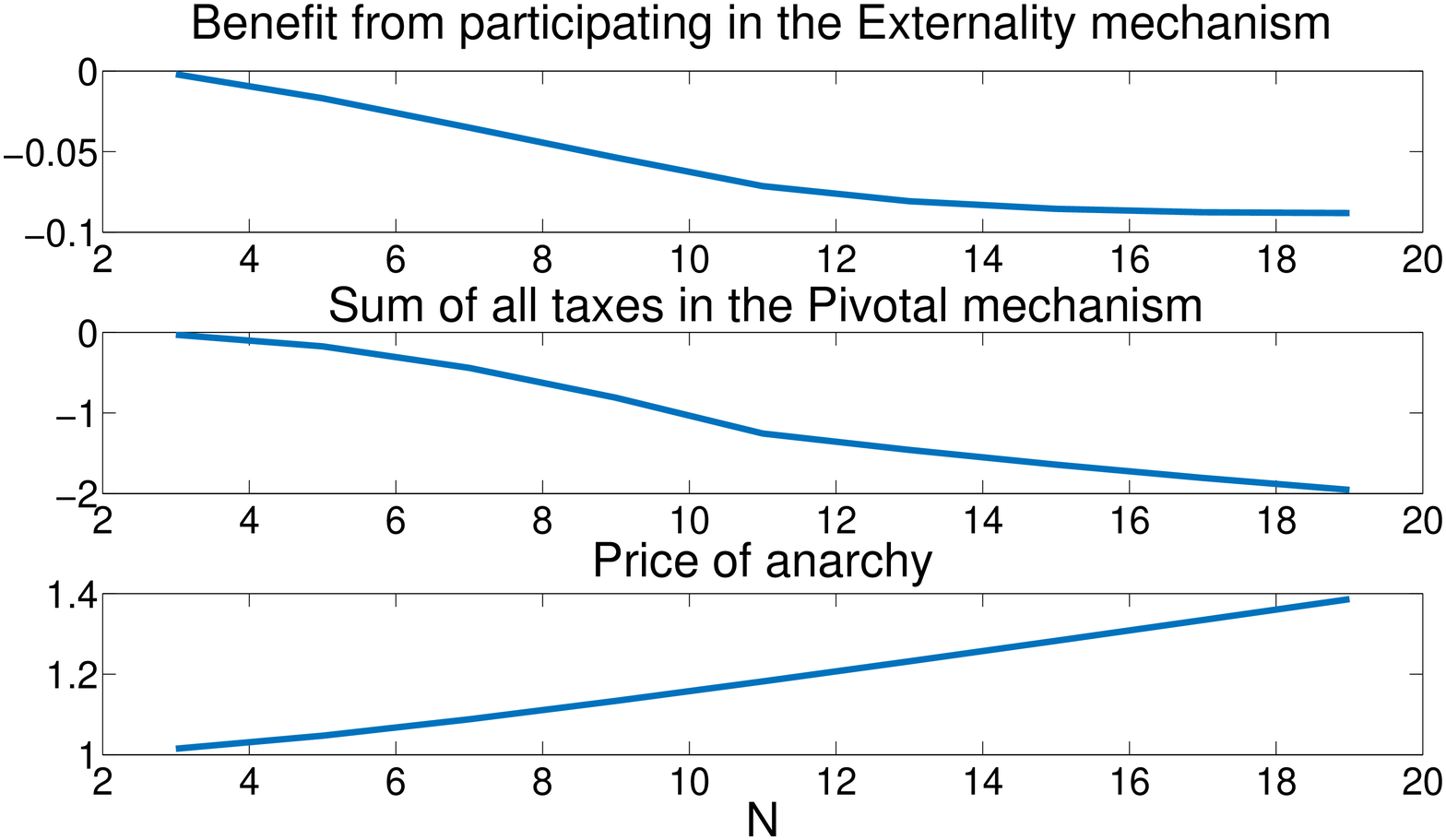}
\caption{Effect of increasing the number of agents $N$}
\label{plot_3_N}
\end{figure}

\paragraph{Changing $N$}
Finally, we set $a=6$ and $c=1$, and increase the number of users $N$, starting from $N=3$. As a result, we will initially be in  [Case $\beta$]: $\hat{x}^i_i>0, \hat{x}^i_j>0$, and gradually move to [Case $\alpha$]: $\hat{x}^i_i=0, \hat{x}^i_j>0$. That is, once enough users participate in the mechanism, the externality is high enough for outliers to stop exerting effort. Figure \ref{plot_3_N} illustrates the results. 

%\begin{figure}
%\centering
%\includegraphics[width=0.7\textwidth]{Plots/Plot_3}
%\caption{Effect of increasing the number of agents $N$}
%\label{plot_3_N}
%\end{figure} 

%%
%\begin{minipage}[t]{\textwidth}
%\centering
%\includegraphics[width=0.5\textwidth]{Plots/Plot_3}
%\caption{Increasing the number of agents $N$}
%\label{plot_3_N}
%\end{minipage}
%\end{figure} 

\subsubsection{Intuitive explanation}
We conclude that as predicted by the analysis, the Pivotal mechanism will always carry a deficit, while the Externality mechanism will always fail to guarantee voluntary participation. We also observe that when the performance of the mechanisms starts improving, the price of anarchy is decreasing, i.e., the reduction in costs from introducing an incentive mechanism is decreasing. This is because the performance of the mechanisms only improves when the system is less interdependent: higher self-dependence, smaller unit costs, or small number of users, all lead to closer to optimal investments by individual users in the state of anarchy. Such users would require smaller incentives to move to the optimal state, hence the reduced budget deficit or smaller participation gap in the Pivotal and Externality mechanisms, respectively. 
We conclude that in these games, when incentive mechanism are more effective, there will be a need for more substantive secondary incentives, or a higher initial budget injection, in order to incentivize optimal investments. 

%%%%%%%%%%%%%%%%%%%%%%
%%%%%%%%%%%%%%%%%%%%%%
%%%%%%%%%%%%%%%%%%%%%%
%%%%%%%%%%%%%%%%%%%%%%
\subsection{Effects of diversity: two classes of self-dependence} \label{sec:sim_2a}
Next, consider a collection of $N$ users with the following total cost functions: 
\[\mathrm{g}_i(\mathbf{x}, t_i) = \exp(-a_{ii}x_i - \sum_{j\neq i} x_j) + c x_i + t_i~.\] 
Assume that we have two classes of users: the self-dependent users $N_1$ for whom $a_{ii} = a_1, i \in N_1$, and the reliant users $N_2=N-N_1$ for whom $a_{ii} = a_2, i \in N_2$. 
We let $c<a_2<1<a_1$. The assumption of $a_2<1<a_1$ entails that users in $N_2$ are affected primarily by other users' security decisions, while those in $N_1$ are more self-dependent. The assumption of $c<a_2<a_1$ ensures that in any equilibrium of the game, at least one user will be exerting non-zero effort. 
 
Similar to the previous section, the socially optimal investment profile and the exit equilibria can be determined according to the first order conditions on users' cost minimization problems subject to non-negative investments. Denote the investments of users in $N_1$ and $N_2$ by $x_1$ and $x_2$, respectively, 
It is easy to show that given the same unit investment costs $c$, and the fact that $a_2<1<a_1$, we get  $x_2=0$ in the socially optimal investment profile. In other words, the users in $N_2$ will never invest in security as they will instead rely on the externality from users in $N_1$. Also, with $c<a_1$, we get $x_1>0$. Therefore, self-dependent users in $N_1$ will be main investors, while reliant users in $N_2$ are free-riders. 
We again omit full analysis for the derivation of the socially optimal solution and the exit equilibria (see online appendix), and limit our discussion to some interesting features of the possible exit equilibria. 

First, it is easy to show that any reliant user $N_2$ staying in the mechanism following the unilateral deviation of one other user will continue as a free-rider. However, when such user unilaterally exits a mechanism, although there always exists an exit equilibrium in which this user continues as a free-rider, there may also exist equilibria under which this user exerts non-zero effort while all other users free-ride. In particular, for $i\in N_2$, an exit equilibrium $\hat{x}^i_i>0, \hat{x}^i_j=0, \forall j\neq i$ exists if and only if: 
\[\frac{a_1+N-2}{c}\leq (\frac{a_2}{c})^{\frac{1}{a_2}}~.\] 
Intuitively, with $a_2$ small enough, given that no other user is investing in security, user $i\in N_2$ will need to exert relatively high effort to reduce its own risk. The considerable externality from this high effort ensures that not investing is a best response for the remaining users. 

Similarly, a user from $N_1$ who leaves the mechanism may continue investing, or become a free-rider. In particular, for $i\in N_1$, an exit equilibrium $\hat{x}^i_i=0, \hat{x}^i_j>0, \forall j\in N_1\backslash\{i\}, \hat{x}^i_k=0, \forall k\in N_2$, where $i\in N_1$ becomes a free-rider, exists if and only if: 
\[(a_1+N_1-2)(\frac{c}{a_1})^{\frac{a_1-1}{N_1-1}} + N_2 \geq a_1\] 

\subsubsection{Simulations}
Assume first that $a_2$ is relatively small, such that when users from $i\in N_2$ step out, $\hat{x}^i_i>0, \hat{x}^i_j=0, \forall j\neq i$ is a possible exit equilibrium.  We gradually increase $a_1$, so that initially $\hat{x}^i_i=0, \hat{x}^i_j>0, \forall j\in N_1\backslash\{i\}, \hat{x}^i_k=0, \forall k\in N_2$ is a possible EE for users from $N_1$, but past a threshold, $\hat{x}^i_i>0, \hat{x}^i_j>0, \forall j\in N_1\backslash\{i\}, \hat{x}^i_k=0, \forall k\in N_2$ is the realized equilibrium. We look at the $N_1$ and $N_2$ users' benefit (in terms of cost reduction) from participating in the Externality mechanism, the budget of the Pivotal mechanism, and the price of anarchy of the game, defined as the sum of costs at the symmetric NE over the total costs at the SO. 

\begin{figure}
\centering
%\begin{minipage}[t]{0.49\columnwidth}
\includegraphics[width=0.8\textwidth]{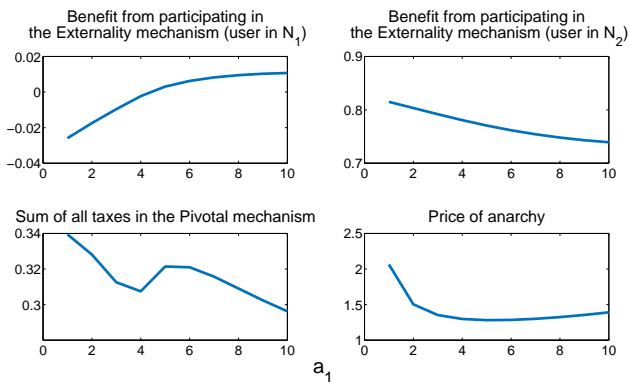}
\caption{Increasing $a_1$ for smaller $a_2$}
\label{small_a2_1}
\end{figure}

In particular, we set $N=10$, and $N_1=8$, $c=0.05$, $a_2=0.1$, and change $a_1\in[1,10]$. The results are depicted in Fig. \ref{small_a2_1}. First, we observe that the Pivotal mechanism will carry a surplus; i.e., WBB holds. Also, VP constraints for users in $N_2$  will be satisfied in the Externality mechanism. However, users from $N_1$ will only have VP when the exit equilibrium is one with $\hat{x}^i_i>0, \hat{x}^i_j>0, \forall j\in N_1\backslash\{i\}, \hat{x}^i_k=0, \forall k\in N_2$. We conclude that these users are only willing to participate in the mechanism if they have to exert non-zero effort even when they stay out. 
%
%\begin{figure}
%\centering
%\includegraphics[width=0.65\textwidth]{Plots/small_a2_large_a1}
%\caption{Effect of increasing $a_1$ for smaller $a_2$}
%\label{small_a2_1}
%\end{figure} 
%

\begin{figure}
\centering
\includegraphics[width=0.8\textwidth]{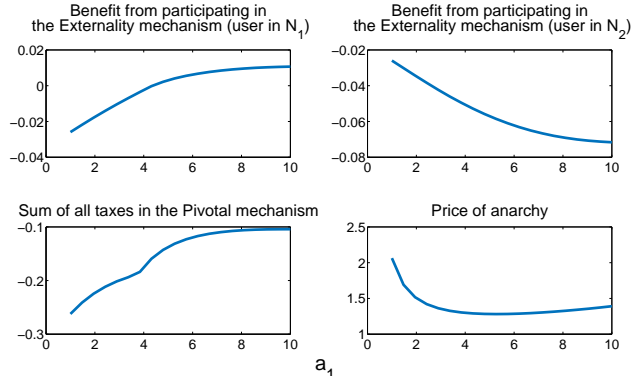}
\caption{Increasing $a_1$ for larger $a_2$}
\label{large_a2_1}
%\end{minipage}
\end{figure}

We next repeat the same simulations, but this time focus on a case with $a_2=0.9$. With this choice of $a_2$, the EE for a user $i\in N_2$ is so that $\hat{x}^i_i=0, \hat{x}^i_j>0, \forall j\in N_1, \hat{x}^i_k=0, \forall k\in N_2\backslash\{i\}$. In other words, these users' will be free-riders whether they participate or not. Consequently, we observe that the participation incentives of users in $N_2$ will no longer be satisfied in the Externality mechanism. In addition, the Pivotal mechanism will carry a budget deficit. These observations are illustrated in Fig. \ref{large_a2_1}. 
%
%\begin{figure}
%\centering
%\includegraphics[width=0.65\textwidth]{Plots/large_a2_large_a1}
%\caption{Effect of increasing $a_1$ for larger $a_2$}
%\label{large_a2_1}
%\end{figure} 
%

%\end{minipage}
%
%\begin{minipage}[t]{0.49\columnwidth}

\subsubsection{Intuitive explanation}
The previous figures illustrate how users' voluntary participation constraints in the Externality mechanism are highly affected by their actions in the exit equilibria. In particular, in the first scenario, we observe that the VP conditions of users in $N_1$ are satisfied only when they are required to exert non-zero effort even when exiting. Similarly, by comparing Figs \ref{small_a2_1} and \ref{large_a2_1}, we see that users in $N_2$ will voluntarily participate if they are forced to act as investors when staying out. Finally, when the exit equilibrium requires users in $N_2$ to invest in security, the Pivotal mechanism is able to extract higher taxes from users in $N_2$, as such equilibrium increases other users' costs considerably compared to the socially optimal solution. This in turn leads to the budget surplus illustrated in Fig. \ref{small_a2_1}.  

%%%%%%%%%%%%%%%%%%%%%%
%%%%%%%%%%%%%%%%%%%%%%
%%%%%%%%%%%%%%%%%%%%%%
%%%%%%%%%%%%%%%%%%%%%%
\subsection{Effects of a single dominant user} \label{sec:sim_dominant}
Consider a collection of $N$ users with weighted total effort risk functions \eqref{eq:u_wte}. Let $a_{1j} = a >1, \forall j$, and $a_{ij}=1, \forall i\neq 1, \forall j$. That is, as $a$ grows, all users' risks become increasingly affected by user $1$'s effort. Thus, users' total cost functions are given by: 
\[\mathrm{g}_i(\mathbf{x}, t_i) = \exp(-ax_1 -\sum_{j=2}^Nx_j) + cx_i + t_i\]
We again assume that $c<1<a$, so as to ensure that at least one user exerts non-zero effort at any equilibrium of the game.

It is easy to show that in a socially optimal investment profile $\mathbf{x}^*$, only user 1 will be exerting effort. This will also be the case when users other than the dominant user leave the mechanism. When the dominant user opts out of the mechanism, however, the exit equilibria will depend on the externality available to this user from the participating nodes. 
The possible equilibria and parameter conditions for which each is possible, as well as the performance of both mechanisms, are summarized in Table \ref{t:vpbb_dominant}. 
%The full analysis is given in {Appendix D}. 
%
\begin{table}
\centering
\caption{Can VP and BB hold simultaneously? -  single dominant user}
\begin{center}
\begin{tabular}{|c|p{3.6cm}|p{2cm}|p{1.6cm}|p{1.2cm}|}
\cline{2-5}
\multicolumn{1}{c|}{} & Exit \newline Equilibrium & Parameter Conditions & VP in \newline Externality & BB in Pivotal\\
\hline
\textbf{CASE $\alpha$} & $\hat{x}^1_1=0, \hat{x}^1_j>0, \forall j\neq 1$ \newline $\hat{x}^i_1>0, \hat{x}^i_j=0 , \forall i,j\neq 1$ & $a<N-1$ & Never & Never \\
\hline
\textbf{CASE $\beta$} & $\hat{x}^1_1>0, \hat{x}^1_j=0, \forall j\neq 1$ \newline $\hat{x}^i_1>0, \hat{x}^i_j=0 , \forall i,j\neq 1$ &$a>N-1$ & Never & Never \\
\hline
\end{tabular}
\end{center}
\label{t:vpbb_dominant}
\end{table}%

\subsubsection{Simulations}
To verify the analysis summarized in Table \ref{t:vpbb_dominant}, we plot a user's benefit from participating in the Externality mechanism (i.e., $g_i(\mathbf{\hat{x}}^i) - \mathrm{g}_i(\mathbf{x}, t_i^E)$), the budget of the Pivotal mechanism (i.e., $\sum_i t_i^P$), and the price of anarchy of the game, as the dependence on the dominant user, $a$, increases. 

\begin{figure}
\centering
\includegraphics[width=0.8\textwidth]{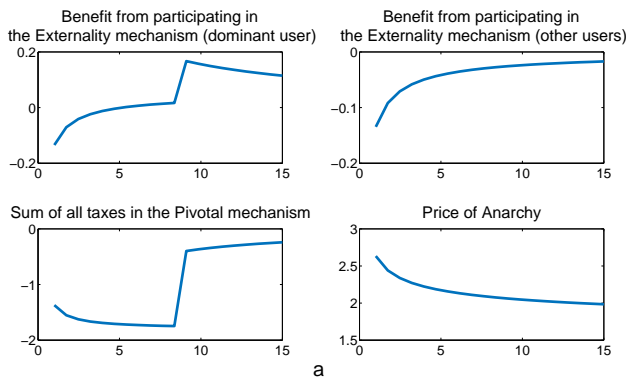}
\caption{Effect of increasing $a$ - single dominant user}
\label{dominant_a}
\end{figure}

In particular, we set $N=10$, $c=0.45$, and increase $a$ from 1 to 15. As a result, we will initially be in {Case $\alpha$}, with $\hat{x}^1_1=0$ and move to {Case $\beta$}, with $\hat{x}^1_1>0$ once $a>N-1$. The results are depicted in Fig. \ref{dominant_a}. As predicted by our analysis, the Pivotal mechanism will always carry a deficit. Also, the voluntary participation condition for non-dominant users will fail under both mechanisms. 

%\begin{figure}
%\centering
%\begin{minipage}[t]{0.24\textwidth}
%\includegraphics[width=\textwidth]{Plots/dom_a_1}
%\end{minipage}
%~
%\begin{minipage}[t]{0.24\textwidth}
%\includegraphics[scale=0.2]{Plots/dom_a_2}
%\end{minipage}
%~
%\begin{minipage}[t]{0.24\textwidth}
%\includegraphics[width=\textwidth]{Plots/dom_a_3}
%\end{minipage}
%~
%\begin{minipage}[t]{0.24\textwidth}
%\includegraphics[width=\textwidth]{Plots/dom_a_4}
%\end{minipage}
%\caption{Effect of increasing $a$ - single dominant user}
%\label{dominant_a}
%\end{figure} 

\subsubsection{Intuitive explanation}
We observe that in these family of games, having a less beneficial equilibrium leads to the voluntary participation of the dominant user, as seen in the top plot in Fig. \ref{dominant_a}. As the exit equilibria for the non-dominant users remains unchanged, so does their participation incentives. Furthermore, we see that no equilibrium can lead to budget surplus in the Pivotal mechanism. That is, although the Pivotal mechanism needs to give out a smaller reward to the dominant user in Case $\beta$ as compared to Case $\alpha$ (hence the jump in the third plot in Fig. \ref{dominant_a}), it still fails to avoid a deficit in both cases, due to the small willingness of free-riders to pay the taxes required to cover this reward. 

%%%%%%%%%%%%%%%%%%%%%%
%%%%%%%%%%%%%%%%%%%%%%
%%%%%%%%%%%%%%%%%%%%%%
%%%%%%%%%%%%%%%%%%%%%%
\subsection{Insights from the weighted total effort model}

First, note that we have identified families of \emph{positive instances}; i.e, problem parameters under which one or both mechanisms can achieve participation and maintain a balanced budget simultaneously. These include Cases $\omega$ and $\zeta$ in Table \ref{t:vpbb_adiag}, which are positive instances for both mechanisms, as well as the region with small $a_1$ and $a_2$ parameters, Fig. \ref{small_a2_1}, which is a positive instance for the Pivotal mechanism. It is also worth mentioning the insight behind the existence of each positive instance:  
\begin{itemize}
\item[-] In Cases $\omega$ and $\zeta$ of Table \ref{t:vpbb_adiag}, incentive mechanisms allow an  \emph{exchange of favors} among users: as all users are mainly dependent on others' investments, they coordinate to each increase their investments in return for improved investments by other users. 
\item[-] In the region with small $a_1$ and $a_2$ parameters in Fig. \ref{small_a2_1}, the Pivotal mechanism is successful as it facilitates \emph{the transfer of funds} from the reliant users to the self-dependent users in return for their improved investments. 
\end{itemize}

Second, we observe that when possible, the selection of exit equilibria that are less beneficial to the outliers helps the performance of both mechanisms. A less beneficial equilibrium can be one that requires a free-rider to become an investor when leaving the mechanism, or one that requires an investor to continue exerting effort when out (although possibly at a lower level). One instance of this feature can be seen by comparing Cases $\omega$ and $\zeta$ with Case $\gamma$ in Table \ref{t:vpbb_adiag}. The same can be observed from Fig. \ref{small_a2_1}, when $a_1$ grows, and also by comparing Figs. \ref{small_a2_1} and \ref{large_a2_1}. Based on this observation, we can expect that in a repeated game setup of security games, by punishing outliers with an appropriate selection of less beneficial equilibria, social optimality, voluntary participation, and budget balance conditions can be simultaneously guaranteed.

\section{Related Work} \label{sec:related} 

The problem of incentivizing optimal security investments in an interconnected system is one example of problems concerning the provision of non-excludable public goods in social and economic networks. Other examples include creation of new parks or libraries at neighborhood level in cities \cite{ballester10}, reducing pollution by neighboring towns \cite{elliott13}, or spread of innovation and research in industry \cite{bramoulle07}. We summarize some of the work most relevant to the current paper. 

\cite{bramoulle07} introduces a network model of public goods, and studies different features of its Nash equilibria. This model is equivalent to a total effort game with linear investment costs and a general interdependence graph. The authors show that these games always have a \emph{specialized} Nash equilibrium; i.e., one in which users are either specialists exerting full effort (equivalent to main investors in our terminology), or free-riders. They show that such equilibria correspond to maximal independent sets of the graph, and that specialized equilibira may lead to higher welfare compared to other (distributed) Nash equilibria.  
Similarly, \cite{ballester06} studies the Nash equilibrium of a linear quadratic interdependence model, and relates the equilibrium effort levels to the nodes' Bonacich centrality in a suitably defined matrix of local complementarities. 
%They also provide a method for identifying the \emph{key player}; i.e., one with the highest effect on the aggregate outcome, using a geometric measure related to the network structure. 
The work in \cite{ballester10} generalizes these results by studying existence, uniqueness, and closed form of the Nash equilibrium in a broader class of games for which best-responses are linear in other players' actions.  All the aforementioned work focuses on the Nash equilibrium in public good provision environments. 

The work of \cite{elliott13} is the most relevant to our work, as it focuses on implementation of Pareto efficient public good outcomes, rather than the Nash equilibria on a given network. The authors define a \emph{benefits matrix} for any given network graph; an entry $B_{ij}$ of the matrix is the marginal rate at which $i$'s effort can be substituted by the externality of $j$'s action. The main result of the paper states that efforts at a Lindahl outcome constitute an eigenvalue centrality vector of this benefits matrix. One such Pareto efficient outcome, the socially optimal outcome, can be implemented using Lindahl taxes determined through the Externality mechanism. The current paper differs from \cite{elliott13} in both modeling and results. First, while a user's action in \cite{elliott13} is strictly costly for the user itself, users in our framework benefit from their own investments as well. More importantly, the focus of \cite{elliott13} is on characterizing users' effort levels in terms of network structure for Lindahl outcomes, the individual rationality of which is established by comparing the Pareto efficient outcome with the state of anarchy, rather than considering unilateral deviations from the mechanism. Our work on the other hand considers both Lindahl and Pivotal taxes, and focuses on users' voluntary participation incentives when unilaterally opting out, as well as tax balance issues.

Finally, in the context of security games, our work in Section \ref{sec:sim} is most related to \cite{walrand11, miura08}. The weighted total effort risk model is a generalization of the total effort model in \cite{varian04}, and is similar to the effective investment model in \cite{walrand11} and the linear influence network game in \cite{miura08}. The linear influence models in \cite{miura08} have been proposed to study properties of the interdependence matrix affecting the existence and uniqueness of the Nash equilibrium. The effective investment model in \cite{walrand11} has been considered to determine a bound on the price of anarchy gap, i.e. the gap between the socially optimal and Nash equilibrium investments, in security games. Our work on the above model complements this literature, by considering the effect of users' interdependence on the performance of incentive mechanisms.

\section{Conclusion} \label{sec:conclusion}

We have shown that in the problem of provision of non-excludable public goods on networks, under general assumptions on the graph structure and users' preferences, it is not possible to design a tax-based incentive mechanism to implement the socially optimal solution while guaranteeing voluntary participation and maintaining a (weakly) balanced budget. Even under a fully connected graph and users with weighted total effort risk functions, we need further conditions on problem parameters (e.g. number of users, the level of interdependence, and cost of investment) to ensure that the well-known Pivotal and Externality mechanisms can achieve social optimality, budget balance, and voluntary participation, simultaneously. These positive instances occur when users can exchange favors by agreeing on increasing their investments, or when they can transfer funds to the more influential users in return for their increased efforts. A comprehensive characterization of problem instances in which all requirements can be simultaneously satisfied remains a main direction of future work.

\section*{Acknowledgment}
The authors would like to thank Armin Sarabi and Hamidreza Tavafoghi for many useful discussions. 
This work is supported by the Department of Homeland Security (DHS) Science and Technology Directorate, Homeland Security Advanced Research Projects Agency (HSARPA), Cyber Security Division (DHS S\&T/HSARPA/CSD), BAA 11-02 via contract number HSHQDC-13-C-B0015. 

\section*{References}
\bibliographystyle{elsarticle-num}

\begin{thebibliography}{10}
\expandafter\ifx\csname url\endcsname\relax
  \def\url#1{\texttt{#1}}\fi
\expandafter\ifx\csname urlprefix\endcsname\relax\def\urlprefix{URL }\fi
\expandafter\ifx\csname href\endcsname\relax
  \def\href#1#2{#2} \def\path#1{#1}\fi

\bibitem{laszka12}
A.~Laszka, M.~Felegyhazi, L.~Butty{\'a}n, A survey of interdependent security
  games, CrySyS 2.

\bibitem{mas-micro}
A.~Mas-Colell, M.~D. Whinston, J.~R. Green, et~al., Microeconomic theory,
  Vol.~1, Oxford university press New York, 1995.

\bibitem{grossklags10}
J.~Grossklags, S.~Radosavac, A.~A. C{\'a}rdenas, J.~Chuang, Nudge:
  Intermediaries' role in interdependent network security, in: Trust and
  Trustworthy Computing, Springer, 2010, pp. 323--336.

\bibitem{bohme10}
R.~B{\"o}hme, G.~Schwartz, Modeling cyber-insurance: Towards a unifying
  framework., in: Workshop on the Economics of Information Security (WEIS),
  2010.

\bibitem{walrand11}
L.~Jiang, V.~Anantharam, J.~Walrand, How bad are selfish investments in network
  security?, IEEE/ACM Transactions on Networking 19~(2) (2011) 549--560.

\bibitem{kun03}
H.~Kunreuther, G.~Heal, Interdependent security, Journal of Risk and
  Uncertainty 26~(2-3) (2003) 231--249.

\bibitem{varian04}
H.~Varian, System reliability and free riding, Economics of information
  security (2004) 1--15.

\bibitem{clarke71}
E.~H. Clarke, Multipart pricing of public goods, Public choice 11~(1) (1971)
  17--33.

\bibitem{hurwicz79}
L.~Hurwicz, Outcome functions yielding walrasian and lindahl allocations at
  nash equilibrium points, The Review of Economic Studies (1979) 217--225.

\bibitem{parkes01}
D.~C. Parkes, Iterative combinatorial auctions: Achieving economic and
  computational efficiency, Ph.D. thesis, University of Pennsylvania (2001).

\bibitem{sharma11}
S.~Sharma, D.~Teneketzis, A game-theoretic approach to decentralized optimal
  power allocation for cellular networks, Telecommunication systems 47~(1-2)
  (2011) 65--80.

\bibitem{grossklags08}
J.~Grossklags, N.~Christin, J.~Chuang, Secure or insure?: a game-theoretic
  analysis of information security games, in: Proceedings of the 17th
  international conference on World Wide Web, ACM, 2008, pp. 209--218.

\bibitem{naghizadeh14b}
P.~Naghizadeh, M.~Liu, Budget balance or voluntary participation? incentivizing
  investments in interdependent security games, in: 52th Annual Allerton
  Conference on Communication, Control, and Computing, IEEE, 2014.

\bibitem{miura08}
R.~Miura-Ko, B.~Yolken, J.~Mitchell, N.~Bambos, Security decision-making among
  interdependent organizations, in: Computer Security Foundations Symposium,
  2008. CSF'08. IEEE 21st, IEEE, 2008, pp. 66--80.

\bibitem{ballester10}
C.~Ballester, A.~Calv{\'o}-Armengol, Interactions with hidden
  complementarities, Regional Science and Urban Economics 40~(6) (2010)
  397--406.

\bibitem{elliott13}
M.~Elliott, B.~Golub, A network approach to public goods, in: Proceedings of
  the fourteenth ACM conference on Electronic commerce, ACM, 2013, pp.
  377--378.

\bibitem{bramoulle07}
Y.~Bramoull{\'e}, R.~Kranton, Public goods in networks, Journal of Economic
  Theory 135~(1) (2007) 478--494.

\bibitem{ballester06}
C.~Ballester, A.~Calv{\'o}-Armengol, Y.~Zenou, Who's who in networks. wanted:
  the key player, Econometrica 74~(5) (2006) 1403--1417.

\end{thebibliography}

\newpage
\section*{Appendix}
\setcounter{section}{0}
%%%%%%%%%%%%%%%%%%%%%%%%%%
%%%%%%%%%%%%%%%%%%%%%%%%%%
%%%%%%%%%%%%%%%%%%%%%%%%%%
%%%%%%%%%%%%%%%%%%%%%%%%%%
\section{Exit equilibria of the weighted total effort model - varying self-dependence} \label{app:adiag}
In this appendix, we solve for the socially optimal investment profile, and identify the possible exit equilibria, and parameter conditions under which each equilibrium is possible. 

The socially optimal investment profile in this game will be given by: 
\[x_i^* = \frac{1}{a+N-1} \ln\frac{a+N-1}{c}, \forall i~.\]
To find the exit equilibrium when a user $i$ steps out, $\mathbf{\hat{x}}^i$, we can write the first order conditions on the users' cost minimization problems. 
To simplify notation, we denote $x:=\hat{x}^i_i$ and $y:=\hat{x}^i_j, \forall j\neq i$. The system of equation determining $x$ and $y$ is given by: 
\begin{align}
-a \exp(-ax-(N-1)y) + c &\geq  0 \notag\\
-(a+N-2) \exp(-x-(a+N-2)y) + c  &\geq 0~.
\label{eq:sys_adiag}  
\end{align} 
There are four possible exit equilibria, depending on the whether $x$ and/or $y$ are non-zero. We look at each case separately. 

\subsection{Exit equilibria with $x>0, y>0$}
Intuitively, when user $i$ steps out, both sides continue to invest in security, perhaps at reduced levels, but no user is fully free-riding. We would need the following to hold simultaneously:  
\begin{align*}
-a \exp(-ax-(N-1)y) + c &= 0 \\
-(a+N-2) \exp(-x-(a+N-2)y) + c &= 0~.
\end{align*}
Let $L_1=\log\frac{a}{c}$ and $L_2 = \log\frac{a+N-2}{c}$. Solving for $x, y$ leads to: 
\begin{align*}
x &= %\frac{(a-1)L_1 - (N-1)(L_2-L_1)}{(a-1)(a+N-1)} = 
\frac{1}{(a-1)(a+N-1)}\log(\frac{a}{c})^{a-1}(1+\frac{N-2}{a})^{-(N-1)}\\
y &= %\frac{(a-1)L_2 + (L_2-L_1)}{(a-1)(a+N-1)} = 
\frac{1}{(a-1)(a+N-1)}\log(\frac{a}{c})^{a-1}(1+\frac{N-2}{a})^{a}~.
\end{align*}
To find the range of parameters for which the above holds, we need to ensure that $x,y$ are indeed positive. 
%First note that $L_2-L_1 = \log\frac{a+N-2}{a}>0$. 
\begin{itemize}
\item 
If $a>1$, then $y>0$. For $x>0$, we need: 
\[{(\frac{a}{c})^{a-1} > (1+\frac{N-2}{a})^{N-1}} \]
\item If $a<1$, then $x>0$. For $y>0$, we need: 
\[{(1+\frac{N-2}{a})^a < (\frac{a}{c})^{1-a}}\]
\end{itemize}

\subsection{Exit equilibria with $x>0, y=0$} 
In this case, the participating users revert to investing zero, so that the outlier is forced to increase its investment:  
\begin{align*}
-a \exp(-ax) + c &= 0 \\
-(a+N-2) \exp(-x) + c &> 0~.
\end{align*}
As a result, we get $x = \frac{1}{a}\log\frac{a}{c}$. For this to be consistent with the second condition, we require: 
\[{(1+\frac{N-2}{a})^a<(\frac{a}{c})^{1-a}}\]
The above always fails to hold for $a>1$, as the LHS is always more than 1, while the RHS is surely less than 1 by the assumption $a>c$. Intuitively, when self-dependence is higher than co-dependence on the outlier, the remaining users will not rely solely on externalities and continue investing when user $i$ steps out. 

For $a<1$ on the other hand, for a small enough $c$ (which in turn leads to higher investment $x$ be the outlier), the equation may hold.

\subsection{Exit equilibria with $x=0, y>0$} 
This means that the loner free-rides, so that we have:  
\begin{align*}
-a \exp(-(N-1)y) + c &> 0 \\
-(a+N-2) \exp(-(a+N-2)y) + c &= 0~.
\end{align*}
As a result, we get $y = \frac{1}{a+N-2}\log\frac{a+N-2}{c}$. For this to be consistent with the first condition, we need: 
\[{(1+\frac{N-2}{a})^{N-1}>(\frac{a}{c})^{a-1}}\]
Note that this always hold for $a<1$, but not necessarily for $a>1$.

\subsection{Exit equilibria with $x=0, y=0$}
We would need the following to hold simultaneously:  
\begin{align*}
-a + c &> 0 \\
-(a+N-2) + c &> 0~, 
\end{align*}
which will never hold, as we initially required that $c<a$.

%%%%%%%%%%%%%%%%%%%%%%%%%%
%%%%%%%%%%%%%%%%%%%%%%%%%%
%%%%%%%%%%%%%%%%%%%%%%%%%%
%%%%%%%%%%%%%%%%%%%%%%%%%%
\section{BB and VP in exit equilibria -  varying self-dependence} \label{app:adiag_bbvp}

In this appendix, we separately analyze each of the possible cases identified in Appendix \ref{app:adiag}, summarized in Table \ref{t:vpbb_adiag}. Specifically, we are interested in the Budget balance condition under the Pivotal mechanism, and users' participation incentives in the Externality mechanism. 

\subsection{Case $\alpha$: fails BB, fails VP} 

In this case, the underlying parameters satisfy $a>1$ and $(1+\frac{N-2}{a})^{N-1}>(\frac{a}{c})^{a-1}$. As a result, the exit equilibrium (EE) is such that $x=0$, and $y= \frac{1}{a+N-2}\log \frac{a+N-2}{c}$. Therefore, the costs of users at the SO and EE are given by: 
\begin{align*}
g_j(\mathbf{x}^*) &= \frac{c}{a+N-1}(1+\log\frac{a+N-1}{c}), \forall j\\
g_j(\mathbf{\hat{x}}^i) &= \frac{c}{a+N-2} (1 + \log{\frac{a+N-2}{c}}), \forall j\neq i\\
g_i(\mathbf{\hat{x}}^i) &= \frac{c}{a+N-2}^{\frac{N-1}{a+N-2}}
\end{align*} 

\paragraph{Budget Balance in the Pivotal mechanism:} 
Note that $\frac{1+\log z}{z}$ is a decreasing function of $z$. Thus, $g_j(\mathbf{\hat{x}}^i)>g_j(\mathbf{x}^*)$ for all $j$, resulting in $t_i^P<0$, indicating rewards to all users $i$, and thus a budget deficit in all scenarios. Intuitively, although when a user $i$ steps out, other users have to invest less in security, thus decreasing their direct investment costs, still their overall security costs go up as a result of the increased risks. Consequently, each user $i$ should be payed a reward to be kept in the mechanism, resulting in a budget deficit. 

\paragraph{Voluntary Participation in the Externality mechanism:} 
Voluntary participation will hold if and only if $g_i(\mathbf{\hat{x}}^i)\geq g_i(\mathbf{x}^*)$, that is: 
\begin{align*}
& \frac{c}{a+N-2}^{\frac{N-1}{a+N-2}} \geq \frac{c}{a+N-1}(1+\log\frac{a+N-1}{c})\\ 
\Leftrightarrow & \frac{c}{a+N-2}^{N-1} \geq (\frac{c}{a+N-1})^{a-1+N-1}(1+\log\frac{a+N-1}{c})^{a-1+N-1}\\
\Leftrightarrow & (\frac{a+N-1}{a+N-2})^{N-1}(1+\frac{N-1}{a})^{a-1}{(\frac{a}{c})}^{a-1} - ({1+\log\frac{a}{c}(1+\frac{N-1}{a})})^{a+N-2}
\geq 0 
%\Leftrightarrow & (\frac{\frac{a}{c}(1+\frac{N-1}{a})}{1+\log\frac{a}{c}(1+\frac{N-1}{a})})^{a-1} 
%\geq (\frac{1+\log\frac{a}{c}(1+\frac{N-1}{a})}{1+\frac{1}{a+N-2}})^{N-1}
\end{align*} 

Based on the last inequality, define the function $g(z):= \kappa_1 z^{a-1} - (1+\log z)^{a+N-2}$. This function is increasing in $z$. 
 As a result, it obtains its maximum when $z$ reaches its maximum value, which by the initial condition is given by $\frac{a}{c} = (1+\frac{N-2}{a})^{\frac{N-1}{a-1}}$. Thus, 
\begin{align*}
g_{max} & = (\frac{a+N-1}{a+N-2})^{N-1}(1+\frac{N-1}{a})^{a-1}(1+\frac{N-2}{a})^{{N-1}} \notag\\
& \qquad - ({1+\log(1+\frac{N-2}{a})^{\frac{N-1}{a-1}}(1+\frac{N-1}{a})})^{a+N-2} \\
& \leq (1+\frac{N-1}{a})^{a+N-2} - ({1+\log(1+\frac{N-1}{a})} +\frac{N-1}{a-1}\log{(1+{\frac{N-2}{a}}}))^{a+N-2}\\
& \leq (1+\frac{N-1}{a})^{a+N-2} - ({1+\log(1+\frac{N-1}{a})} +\frac{N-1}{a}\log{(1+{\frac{N-2}{a}}}))^{a+N-2}
\end{align*} 
Let $z:=\frac{N-1}{a}$, and define $f(z):=\log(1+z)+z\log(1+z-\frac1a) - z$ (i.e., we are assuming a fixed $a$). The derivative of this function wrt $z$ is given by: 
\[\frac{1}{1+z} + \log(1+z - \frac{1}{a}) +  \frac{z}{1+z - \frac{1}{a}} -1 = \log(1+z - \frac{1}{a}) + \frac{\frac1a z}{(1+z)(1-\frac1a +z)}~.\]
As the above is positive for all $a>1$, we conclude that $f(z)$ is an increasing function in $z$. Furthermore, $\lim_{z\rightarrow 0} f(z) = 0$, which in turn means that $f(z)\geq 0, \forall z\geq 0$, and therefore, $g_{max}$ is always non-positive. This in turn means that the VP condition can never be satisfied. 

%It is interesting to consider the following: given that the condition $(1+\frac{N-2}{a})^{N-1}>(\frac{a}{c})^{a-1}$ holds in this case, we have $\frac{c}{a+N-2}^{\frac{N-1}{a+N-2}}< \frac{c}{a}$ (which is intuitively saying that the loner always faces a higher risk when stepping out to this equilibrium. The user however saves the investment costs, and that's why participation incentives have to be addressed). 

\subsection{Case $\beta$: fails BB, fails VP} 

For this case, the underlying parameters satisfy $a>1$ and $(1+\frac{N-2}{a})^{N-1}<(\frac{a}{c})^{a-1}$. As a result, the exit equilibrium (EE) is such that $x>0, y>0$, and are given by $x= \frac{1}{(a-1)(a+N-1)}\log{(\frac{a}{c})^{a-1}}{(1+{\frac{N-2}{a}})^{-(N-1)}}$ and \linebreak $y= \frac{1}{(a-1)(a+N-1)}\log{(\frac{a}{c})^{a-1}}{(1+{\frac{N-2}{a}})^{a}}$. Therefore, the costs of users at the SO and EE are given by: 
\begin{align*}
g_j(\mathbf{x}^*) &= \frac{c}{a+N-1}(1+\log\frac{a+N-1}{c}), \forall j\\
g_j(\mathbf{\hat{x}}^i) &= \frac{c}{a+N-2} + \frac{c}{(a-1)(a+N-1)}\log{(\frac{a}{c})^{a-1}}{(1+{\frac{N-2}{a}})^{a}}, \forall j\neq i\\
g_i(\mathbf{\hat{x}}^i) &= \frac{c}{a} + \frac{c}{(a-1)(a+N-1)}\log{(\frac{a}{c})^{a-1}}{(1+{\frac{N-2}{a}})^{-(N-1)}}
\end{align*} 

\paragraph{Budget Balance in the Pivotal mechanism:} 
For the mechanism to have a budget deficit we would need $g_j(\mathbf{\hat{x}}^i)\geq g_j(\mathbf{x}^*)$, which holds if and only if: 
\begin{align*}
& \frac{c}{a+N-1}(1+\log\frac{a+N-1}{c}) \leq \\
& \hspace{1in} \frac{c}{a+N-2} + \frac{c}{(a-1)(a+N-1)}\log{(\frac{a}{c})^{a-1}}{(1+{\frac{N-2}{a}})^{a}}\\
\Leftrightarrow & 1+\log\frac{a}{c}(1+\frac{N-1}{a}) \leq 1 + \frac{1}{a+N-2} + \log{\frac{a}{c}}{(1+{\frac{N-2}{a}})^{\frac{a}{a-1}}}\\
\Leftrightarrow & \log(1+\frac{N-1}{a}) \leq \frac{1}{a+N-2} + {\frac{a}{a-1}}\log{(1+{\frac{N-2}{a}})}\\
\Leftarrow &  \log(1+\frac{N-1}{a}) \leq \frac{1}{a+N-2} + \log{(1+{\frac{N-2}{a}})}\\
\Leftrightarrow &  \log(1+\frac{1}{a+N-2}) \leq \frac{1}{a+N-2} 
\end{align*} 
The last line is true because $\log(1+x)\leq x$, for all $x>0$. Therefore, the mechanism always carries a budget deficit. 

\paragraph{Voluntary Participation in the Externality mechanism:} 
The mechanism fails voluntary participation if and only if: 
\begin{align*}
& \frac{c}{a+N-1}(1+\log\frac{a+N-1}{c}) \geq \\
& \hspace{1in} \frac{c}{a} + \frac{c}{(a-1)(a+N-1)}\log{(\frac{a}{c})^{a-1}}{(1+{\frac{N-2}{a}})^{-(N-1)}}\\
\Leftrightarrow & 1+\log\frac{a}{c}(1+\frac{N-1}{a}) \geq 1 + \frac{N-1}{a} + \log{\frac{a}{c}}{(1+{\frac{N-2}{a}})^{\frac{-(N-1)}{a-1}}}\\
\Leftrightarrow & \log(1+\frac{N-1}{a})+ {\frac{N-1}{a-1}}\log{(1+{\frac{N-2}{a}})} \geq \frac{N-1}{a}\\
\Leftarrow & \log(1+\frac{N-1}{a})+ {\frac{N-1}{a}}\log{(1+{\frac{N-1}{a}} - \frac1a)} \geq \frac{N-1}{a}\\
\end{align*} 
Let $z:=\frac{N-1}{a}$, and define $f(z):=\log(1+z)+z\log(1+z-\frac1a) - z$ (i.e., we are assuming a fixed $a$). The derivative of this function wrt $z$ is given by: 
\[\frac{1}{1+z} + \log(1+z - \frac{1}{a}) +  \frac{z}{1+z - \frac{1}{a}} -1 = \log(1+z - \frac{1}{a}) + \frac{\frac1a z}{(1+z)(1-\frac1a +z)}~.\]
As the above is positive for all $a>1$, we conclude that $f(z)$ is an increasing function in $z$. Furthermore, $\lim_{z\rightarrow 0} f(z) = 0$, which in turn means that $f(z)\geq 0, \forall z\geq 0$, and therefore, that the VP condition always fails to hold under these parameter settings.

\subsection{Case $\gamma$: fails BB, fails VP} 

Here, we only require that $a<1$, and all other values of $N$ or $c$ will guarantee the existence of an equilibrium $x=0$ and $y= \frac{1}{a+N-2}\log \frac{a+N-2}{c}$. This is thus parallel with Case $\alpha$. Users' costs in the SO and EE are similarly given by:
\begin{align*}
g_j(\mathbf{x}^*) &= \frac{c}{a+N-1}(1+\log\frac{a+N-1}{c}), \forall j\\
g_j(\mathbf{\hat{x}}^i) &= \frac{c}{a+N-2} (1 + \log{\frac{a+N-2}{c}}), \forall j\neq i\\
g_i(\mathbf{\hat{x}}^i) &= \frac{c}{a+N-2}^{\frac{N-1}{a+N-2}}
\end{align*} 

\paragraph{Budget Balance in the Pivotal mechanism:} 
Note that $\frac{1+\log z}{z}$ is a decreasing function of $z$. Thus, $g_j(\mathbf{\hat{x}}^i)>g_j(\mathbf{x}^*)$ for all $j$, resulting in $t_i^P<0$, indicating rewards to all users $i$, and thus a budget deficit in all scenarios (exactly similar to case $\alpha$). 

\paragraph{Voluntary Participation in the Externality mechanism:} 
Voluntary participation will fail if and only if $g_i(\mathbf{\hat{x}}^i)\leq g_i(\mathbf{x}^*)$, that is: 
\begin{align*}
& \frac{c}{a+N-2}^{\frac{N-1}{a+N-2}} \leq \frac{c}{a+N-1}(1+\log\frac{a+N-1}{c})\\ 
\Leftrightarrow & \frac{c}{a+N-2}^{N-1} \leq (\frac{c}{a+N-1})^{a-1+N-1}(1+\log\frac{a+N-1}{c})^{a-1+N-1}\\
\Leftrightarrow & (\frac{\frac{a}{c}(1+\frac{N-1}{a})}{1+\log\frac{a}{c}(1+\frac{N-1}{a})})^{a-1} 
\geq (\frac{1+\log\frac{a}{c}(1+\frac{N-1}{a})}{1+\frac{1}{a+N-2}})^{N-1}
\end{align*} 
First, we note that the RHS is always greater than 1, as $1+\log x \leq x$. On the other hand, since $a<1$, $\frac{1}{a+N-2}<\log\frac{a}{c}(1+\frac{N-1}{a})$ holds for all $N\geq 3$, so that the LHS will be less than 1. Therefore, the VP condition always fails.

\subsection{Case $\zeta$: has BB, has VP} 

This case has equilibrium investments similar to case $\beta$, but under parameter conditions $a<1$, and $(1+\frac{N-2}{a})^a < (\frac{a}{c})^{1-a}$. Therefore, we have the following costs for the users: 
\begin{align*}
g_j(\mathbf{x}^*) &= \frac{c}{a+N-1}(1+\log\frac{a+N-1}{c}), \forall j\\
g_j(\mathbf{\hat{x}}^i) &= \frac{c}{a+N-2} + \frac{c}{(a-1)(a+N-1)}\log{(\frac{a}{c})^{a-1}}{(1+{\frac{N-2}{a}})^{a}}, \forall j\neq i\\
g_i(\mathbf{\hat{x}}^i) &= \frac{c}{a} + \frac{c}{(a-1)(a+N-1)}\log{(\frac{a}{c})^{a-1}}{(1+{\frac{N-2}{a}})^{-(N-1)}}
\end{align*} 

\paragraph{Budget Balance in the Pivotal mechanism:} 
For the mechanism to have budget balance we would need $g_j(\mathbf{\hat{x}}^i)\leq g_j(\mathbf{x}^*)$, which holds if and only if: 
\begin{align*}
& \frac{c}{a+N-1}(1+\log\frac{a+N-1}{c}) \geq \\
& \hspace{1in} \frac{c}{a+N-2} + \frac{c}{(a-1)(a+N-1)}\log{(\frac{a}{c})^{a-1}}{(1+{\frac{N-2}{a}})^{a}}\\
\Leftrightarrow & 1+\log\frac{a}{c}(1+\frac{N-1}{a}) \geq 1 + \frac{1}{a+N-2} + \log{\frac{a}{c}}{(1+{\frac{N-2}{a}})^{\frac{a}{a-1}}}\\
\Leftrightarrow & \log(1+\frac{N-1}{a}) \geq \frac{1}{a+N-2} + {\frac{a}{a-1}}\log{(1+{\frac{N-2}{a}})}\\
\Leftarrow &  \log(1+\frac{N-1}{a}) \geq \frac{1}{a+N-2}
\end{align*} 
The last line follows from the previous because $a<1$, and is true because its LHS is $\geq \log N$ and its RHS is $\leq 1/(N-1)$. Therefore, the mechanism always has budget balance in this scenario. 

\paragraph{Voluntary Participation in the Externality mechanism:} 
The mechanism has voluntary participation if and only if: 
\begin{align*}
& \frac{c}{a+N-1}(1+\log\frac{a+N-1}{c}) \leq \\
& \hspace{1in} \frac{c}{a} + \frac{c}{(a-1)(a+N-1)}\log{(\frac{a}{c})^{a-1}}{(1+{\frac{N-2}{a}})^{-(N-1)}}\\
\Leftrightarrow & 1+\log\frac{a}{c}(1+\frac{N-1}{a}) \leq 1 + \frac{N-1}{a} + \log{\frac{a}{c}}{(1+{\frac{N-2}{a}})^{\frac{-(N-1)}{a-1}}}\\
\Leftrightarrow & \log(1+\frac{N-1}{a}){(1+{\frac{N-2}{a}})^{\frac{N-1}{a-1}}} \leq \frac{N-1}{a}
\end{align*} 
The last statement holds because the second element in the logarithm is always less than 1, due to $a<1$, and the result follows as $\log(1+z)\leq z$, for all $z>0$.

\subsection{Case $\omega$: has BB, has VP} 

The last case in realized under parameter settings $a<1$ and $(1+\frac{N-2}{a})^a < (\frac{a}{c})^{1-a}$, and $x=\log\frac{a}{c}$ and $y=0$ is the possible exit equilibrium. The users' costs in the SO and EE here are given by: 
\begin{align*}
g_j(\mathbf{x}^*) &= \frac{c}{a+N-1}(1+\log\frac{a+N-1}{c}), \forall j\\
g_j(\mathbf{\hat{x}}^i) &= (\frac{c}{a})^{\frac1a},  \forall j\neq i\\
g_i(\mathbf{\hat{x}^i}) &= \frac{c}{a}(1+\log\frac{a}{c})~.
\end{align*}

%First, it is interesting to note that such exit equilibrium is also a Nash equilibrium. This is because user $i$ is already best-responding to the profile $x_j=0, \forall j\neq i$. On the other hand, for any single user $j\neq i$, the first derivative on the cost at this equilibrium is positive: 
%\[-a\exp(-x) + c = -a (\frac{c}{a})^{\frac1a} + c = -a( (\frac{c}{a})^{\frac1a} - \frac{c}{a} )>0~.\]
%Therefore, each individual user $j$'s costs are increasing at this equilibrium, and investing $y=0$ is a best response. 

\paragraph{Budget Balance in the Pivotal mechanism:} 
First we use $(1+\frac{N-2}{a})^a < (\frac{a}{c})^{1-a}$ to conclude that $(\frac{c}{a})^{\frac1a}\leq \frac{c}{a+N-2}$. Now, for the mechanism to have budget balance we would need $g_j(\mathbf{\hat{x}}^i)\leq g_j(\mathbf{x}^*)$, which holds if and only if: 
\begin{align*}
& \frac{c}{a+N-1}(1+\log\frac{a+N-1}{c}) \geq (\frac{c}{a})^{\frac1a}\\
\Leftarrow & 1+\log\frac{a}{c}(1+\frac{N-1}{a}) \geq 1 + \frac{1}{a+N-2}\\
\Leftarrow &  \log(1+\frac{N-1}{a}) \geq \frac{1}{a+N-2}
\end{align*} 
where the last line line follows from the previous because $\frac{a}{c}>1$, and is true because its LHS is $\geq \log N$ and its RHS is $\leq 1/(N-1)$. Therefore, the mechanism always has budget balance in this scenario.

\paragraph{Voluntary Participation in the Externality mechanism:} 
As $\frac{1+\log x}{x}$ is a decreasing function in $x$ when $x>1$, and $1<\frac{a}{c}<\frac{a+N-1}{c}$, the costs when staying out are higher for user $i$. Therefore VP is satisfied in the Externality mechanism in this case.

%%%%%%%%%%%%%%%%%%%%%%%%%%
%%%%%%%%%%%%%%%%%%%%%%%%%%
%%%%%%%%%%%%%%%%%%%%%%%%%%
%%%%%%%%%%%%%%%%%%%%%%%%%%
%\subsection*{Bounds on the SO solution - two self-dependence classes}
%As mentioned, the equation determining the SO solution under two classes of self-dependence does not have a close form solution. However, it is possible to find a lower bound and an upper bound on the solution $x^*_1$. This is because the LHS of the above equation is decreasing in $x_1$. Therefore, we first use $\exp(-x)\geq 1-x$, and obtain the following lower bound on the LHS: 
%\begin{align*}
%(a_1+N_1-1)\exp(-(a_1+N_1-1)x_1) + N_2\exp(-N_1x_1) \geq (a_1+N_1-1)(1-(a_1+N_1-1)x_1) + N_2(1-N_1x_1)~.
%\end{align*} 
%Consequently, $x_1^*\geq \frac{a_1+N-1-c}{(a_1+N_1-1)^2+N_1N_2}$. 
%
%We next use an upper bound for the LHS and obtain a upper bound on the solution. Using the fact that $a_1>1$: 
%\begin{align*}
%(a_1+N_1-1)\exp(-(a_1-1)x_1 - N_1x_1) + N_2\exp(-N_1x_1) \leq (a_1+N-1)\exp(-N_1x_1)~.
%\end{align*} 
%Consequently, $x_1^*\leq \frac{1}{N_1}\log\frac{a_1+N-1}{c}$. 
%
%Put together, we can bound the investments at the socially optimal solution as follows: 
%\[\boxed{\frac{a_1+N-1-c}{(a_1+N_1-1)^2+N_1N_2} \leq x_1^*\leq \frac{1}{N_1}\log\frac{a_1+N-1}{c}}\]

%%%%%%%%%%%%%%%%%%%%%%%%%%
%%%%%%%%%%%%%%%%%%%%%%%%%%
%%%%%%%%%%%%%%%%%%%%%%%%%%
%%%%%%%%%%%%%%%%%%%%%%%%%%
\section{Exit equilibria of the weighted total effort model - two classes of self-dependence} \label{app:twoa}
In this appendix, we present a (partial) analysis of possible exit equilibria, and parameter conditions under which each is possible, for the wighted total effort family described in Section \ref{sec:sim_2a}. 
Denoting the investments of the users in $N_1$ and $N_2$ by $x_1$ and $x_2$, respectively, the socially optimal investment profile in this game is determined according to the first order conditions on users' cost minimization problems:   
\begin{align*}
&-(a_1+N_1-1)\exp(-(a_1+N_1-1)x_1 - N_2x_2) - N_2\exp(-N_1x_1 - (a_2+N_2-1)x_2) + c \geq 0~,\\
&-N_1\exp(-(a_1+N_1-1)x_1 - N_2x_2) - (a_2+N_2-1)\exp(-N_1x_1 - (a_2+N_2-1)x_2) + c \geq 0~.
\end{align*}
It is easy to see that at this socially optimal solution, $x_2=0$, and $x_1$ is given by: 
\begin{align*}
(a_1+N_1-1)\exp(-(a_1+N_1-1)x_1) + N_2\exp(-N_1x_1) = c~.
\end{align*} 
In general, the above equation does not have a closed form solution. However, it is possible to find a lower bound and an upper bound on the solution $x^*_1$. It is also possible to solve for $x^*_1$ numerically. 
Once we solve for this socially optimal investment, we can determine the taxes assigned by the Externality mechanism: 
\begin{align*}
t_i^E &= (a_1+N_1-1)x_1\exp(-(a_1+N_1-1)x_1) - cx_1, &\quad \forall i\in N_1~,\\
t_i^E &= N_1x_1\exp(-N_1x_1), &\quad \forall i\in N_2~. 
\end{align*}
Note that the sum of the above taxes is indeed zero. Also, it is interesting to note that as expected, the free-riders in $N_2$ always pay a tax, while the main investors in $N_1$ receive a subsidy.  
In order to find exit equilibria in this family of games, we would again need to solve equations with a similar format to that of the socially optimal solution, which in general lack a closed form solution. We therefore do not include a full analysis of this scenario, and limit our discussion to some interesting features of the possible exit equilibria.

\subsection{{Exit equilibrium: a user from $N_1$ leaving}} 
Let $x$ denote the investment of the deviating user from group $N_1$, and $y_1$ and $y_2$ denote the investments of users remaining in $N_1$ and $N_2$. We have the following system of equations: 
\begin{align*}
&-a_1\exp(-a_1x-(N_1-1)y_1-N_2y_2) + c &\geq 0~,\\
&-(a_1+N_1-2)\exp(-x-(a_1+N_1-2)y_1 - N_2y_2) \\
& \hspace{0.5in} - N_2\exp(-x-(N_1-1)y_1 - (a_2+N_2-1)y_2) + c &\geq 0~,\\
&-(N_1-1)\exp(-x-(a_1+N_1-2)y_1 - N_2y_2) \\
& \hspace{0.5in} - (a_2+N_2-1)\exp(-x-(N_1-1)y_1 - (a_2+N_2-1)y_2) + c &\geq 0~.
\end{align*} 
By an argument similar to that in the derivation of the SO solution in \ref{sec:sim_2a}, we will always have $y_2=0$. As a result, the system of equations reduces to: 
\begin{align*}
&-a_1\exp(-a_1x-(N_1-1)y_1) + & c \geq 0~,\\
&-(a_1+N_1-2)\exp(-x-(a_1+N_1-2)y_1) - N_2\exp(-x-(N_1-1)y_1) + & c \geq 0~.
\end{align*} 
We consider the following possible cases, depending on whether $x$ and/or $y_1$ are non-zero. 

\subsubsection{Exit equilibria with $x>0, y_1>0$} 
This would require a solution to the following system of equations: 
\begin{align*}
&a_1\exp(-a_1x-(N_1-1)y_1) &= c~,\\
&(a_1+N_1-2)\exp(-x-(a_1+N_1-2)y_1) + N_2\exp(-x-(N_1-1)y_1) &= c~.
\end{align*}
%We can bound the solution as follows: 
%\[x^*+(N_1-1)y_1^* \leq \log\frac{a_1+N-2}{c}~,\]
%and, 
%\[{a_1+N-2-c}\leq (a_1+N-2)x^*+((a_1+N_1-2)^2+(N_1-1)N_2)y_1^*~. \]
%Given the first equation, the first inequality is trivial. 
The above equations tell us that: 
\[a_1x+(N_1-1)y_1 = \log\frac{a_1}{c}~.\]
We can now substitute $y_1$ in the second equation to find $x$:
\begin{align*}
&(a_1+N_1-2)\exp(-[\log\frac{a_1}{c}-(a_1-1)x] \\
&- \frac{a_1-1}{N_1-1}[\log\frac{a_1}{c}-a_1x]) + N_2\exp(-[\log\frac{a_1}{c}-(a_1-1)x]) = c~, 
\end{align*}
which can be solved numerically. 
 
\subsubsection{Exit equilibria with $x=0, y_1>0$} 
This would require: 
\begin{align*}
&a_1\exp(-(N_1-1)y_1) &\leq c~,\\
&(a_1+N_1-2)\exp(-(a_1+N_1-2)y_1) + N_2\exp(-(N_1-1)y_1) &= c~.
\end{align*}
%We can bound the solution to the second equation, which determins $y_1$, as follows: 
%\[{\frac{a_1+N-2-c}{(a_1+N_1-2)^2+(N_1-1)N_2} \leq y_1^*\leq \frac{1}{N_1-1}\log\frac{a_1+N-2}{c}}\]
The first equation above can be used to find a lower-bound on $y_1$. Intuitively, the investments $y_1$ should be high enough for the outlier to decide against investing (i.e., set $x=0$). We have: 
\[y_1 \geq \frac{1}{N_1-1}\log\frac{a_1}{c}~.\]
Given that the LHS of the second equation, which determins $y_1$, is decreasing in $y_1$, the above system of equation is consistent if and only if: 
\[{(a_1+N_1-2)(\frac{c}{a_1})^{\frac{a_1-1}{N_1-1}} + N_2 \geq a_1~.}\] 

\subsubsection{Exit equilibria with $x>0, y_1=0$} 
This requires that $x=\frac{1}{a_1}\log\frac{a_1}{c}$, and that: 
\begin{align*}
& -(a_1+N_1-2)\exp(-x) - N_2\exp(-x) +  c \geq 0 \\
\Leftrightarrow & (a_1+N-2)(\frac{c}{a_1})^{\frac{1}{a_1}} \leq c \Leftrightarrow (1+\frac{N-2}{a_1})^{a_1}\leq (\frac{c}{a_1})^{{a_1}-1}~. 
\end{align*}
It is easy to see that the LHS is always greater than 1, while the RHS is less than 1, making this exit equilibrium impossible. 

\subsubsection{Exit equilibria with $x=0, y_1=0$} 
This case will never happen, as $c<a_1$.

\subsection{{Exit equilibrium: a user from $N_2$ leaving}} 
Let $x$ denote the investment of the deviating user from group $N_2$, and $y_1$ and $y_2$ denote the investments of users remaining in $N_1$ and $N_2$. We have the following system of equations: 
\begin{align*}
&-a_2\exp(-a_2x-N_1y_1-(N_2-1)y_2) &+ c \geq 0~,\\
&-(a_1+N_1-1)\exp(-x-(a_1+N_1-1)y_1 - (N_2-1)y_2) \\
& \hspace{0.5in} - (N_2-1)\exp(-x-N_1y_1 - (a_2+N_2-2)y_2) &+ c \geq 0~,\\
&-N_1\exp(-x-(a_1+N_1-1)y_1 - (N_2-1)y_2) \\
& \hspace{0.5in} - (a_2+N_2-2)\exp(-x-N_1y_1 - (a_2+N_2-2)y_2) &+ c \geq 0~.
\end{align*} 
By an argument similar to that in the derivation of the SO solution in \ref{sec:sim_2a}, we will always have $y_2=0$. As a result, the system of equations reduces to: 
\begin{align*}
&-a_2\exp(-a_2x-N_1y_1) + c &\geq 0~,\\
&-(a_1+N_1-1)\exp(-x-(a_1+N_1-1)y_1) - (N_2-1)\exp(-x-N_1y_1) + c &\geq 0~. 
\end{align*} 
We consider the following possible cases. 

\subsubsection{Exit equilibria with $x>0, y_1>0$} 
This would require a solution to the following system of equations: 
\begin{align*}
&a_2\exp(-a_2x-N_1y_1) &= c~,\\
&(a_1+N_1-1)\exp(-x-(a_1+N_1-1)y_1) + (N_2-1)\exp(-x-N_1y_1) &= c~. 
\end{align*}
From the first equation we know that $a_2x+N_1y_1 = \log\frac{a_2}{c}$. 
%Now, we first upper bound the solution of the second equation: 
%\[x^*+N_1y_1^* \leq \log\frac{a_1+N-2}{c}~.\]
%Combining this with the conclusion from the first equation, we conclude that this would require: 
%\[(1-a_2)x^* \leq \log\frac{a_1+N-2}{a_2}~.\]
To solve the system, we substitute $y$ in the second equation and obtain: 
\begin{align*}
&(a_1+N_1-1)\exp(-\log\frac{a_2}{c}-(1-a_2)x - \frac{a_1-1}{N_1}[\log\frac{a_2}{c}-a_2x]) \\
& \hspace{1in} + (N_2-1)\exp(-\log\frac{a_2}{c}-(1-a_2)x) = c~,  
\end{align*} 
which can be solved numerically. 

\subsubsection{Exit equilibria with $x=0, y_1>0$} 
This would require: 
\begin{align*}
&a_2\exp(-N_1y_1) &\leq c~,\\
&(a_1+N_1-1)\exp(-(a_1+N_1-1)y_1) + (N_2-1)\exp(-N_1y_1) &= c~. 
\end{align*}
Note that from the equation above we can say:
\[\exp(-N_1y_1) = \frac{c-(a_1+N_1-1)\exp(-(a_1+N_1-1)y_1)}{N_2-1}\leq c \leq \frac{c}{a_2}~.\]
Therefore, this is always a possible equilibrium. 

\subsubsection{Exit equilibria with $x>0, y=0$} 
This requires that $x=\frac{1}{a_2}\log\frac{a_2}{c}$, and that: 
\begin{align*}
& -(a_1+N_1-1)\exp(-x) - (N_2-1)\exp(-x) +  c \geq 0 \\
& \Leftrightarrow (a_1+N-2)(\frac{c}{a_2})^{\frac{1}{a_2}} \leq c \\
& \Leftrightarrow \frac{a_1+N-2}{c}\leq (\frac{a_2}{c})^{\frac{1}{a_2}}~. 
\end{align*}
The above holds when $a_2$ is small (RHS is maximized at $a_2=c\exp(1)$). 

\subsubsection{Exit equilibria with $x=0, y=0$} 
This case will never happen, as $c<a_2$.

%%%%%%%%%%%%%%%%%%%%%%%%%%
%%%%%%%%%%%%%%%%%%%%%%%%%%
%%%%%%%%%%%%%%%%%%%%%%%%%%
%%%%%%%%%%%%%%%%%%%%%%%%%%
\section{Exit equilibria of the weighted total effort model - single dominant user} \label{app:dominant}
In this appendix, we solve for the socially optimal investment profile, and identify the possible exit equilibria, and parameter conditions under which each equilibrium is possible. 
It is easy to show that in a socially optimal investment profile $\mathbf{x}^*$, only user 1 will be exerting effort, so that: 
\[x_1^* = \frac{1}{a}\ln\frac{aN}{c}, \quad x_j^*=0, \forall j=2, \ldots, N~.\]
We next find the exit equilibria under two different cases. First, if any non-dominant user $i\neq 1$ steps out of the mechanism, user $1$ will continue exerting all effort, but at a lower level given by: 
\[\hat{x}^i_1 = \frac{1}{a}\log\frac{a(N-1)}{c}, \quad \hat{x}^i_j=0, \forall j=2, \ldots, N~.\]
Next, if user 1 steps out of the mechanism, there are two possible exit equilibria: if $a>N-1$, there will be enough externality for users $j\neq 1$ to continue free-riding, resulting in the following equilibrium investment levels: 
\[\hat{x}^1_1 = \frac{1}{a}\log\frac{a}{c}, \quad \hat{x}^1_j=0, \forall j=2, \ldots, N~.\]
However, when $a<N-1$, user 1 will free-ride on the externality of other users' investments, leading to the exit equilibrium: 
\[\hat{x}^1_1 = 0, \quad \hat{x}^1_j=\frac{1}{N-1}\log\frac{N-1}{c}, \forall j=2, \ldots, N~.\]

%%%%%%%%%%%%%%%%%%%%%%%%%%
%%%%%%%%%%%%%%%%%%%%%%%%%%
%%%%%%%%%%%%%%%%%%%%%%%%%%
%%%%%%%%%%%%%%%%%%%%%%%%%%
\section{BB and VP in exit equilibria -  single dominant user} \label{app:dominant_bbvp}
In this appendix, we look at the performance of the Pivotal and Externality mechanisms, under the different exit equilibria identified in Section \ref{sec:sim_dominant}, summarized in Table \ref{t:vpbb_dominant}. 

In the Externality mechanism, users' taxes are given by: 
\begin{align*}
t_1^E(\mathbf{x}^*) &= cx_1^*(\frac{1}{N}-1)\\
t_j^E(\mathbf{x}^*) &=\frac{c}{N}x_1^*, \forall j=2, \ldots, N~. 
\end{align*}
For non-dominant users $i\in\{2, \ldots, N\}$ to voluntarily participate in the mechanism, we require $g_i(\mathbf{\hat{x}}^i) \geq \mathrm{g}_i(\mathbf{x}^*, t_i^E(\mathbf{x}^*))$: 
\begin{align*}
&\frac{c}{a(N-1)} \geq \frac{c}{aN} + \frac{c}{aN}\log\frac{aN}{c}\\
\Leftrightarrow& \frac{1}{N-1} \geq \log{N}+\log\frac{a}{c}~.
\end{align*}
However, $\log{N}\geq \frac{1}{N-1}, \forall N\geq 3$, and $a>c$. Therefore, the voluntary participation constraints will always fail to hold in the Externality mechanism. 

Finally, we analyze the total budget in the Pivotal mechanism for the current scenario. The taxes for the non-dominant users $i\neq 1$ will be given by: 
\[t_i^P = (N-1)\frac{c}{aN} + cx_1^* - (N-1)\frac{c}{a(N-1)} - c\hat{x}^i_1 = \frac{c}{a}(\log\frac{N}{N-1} - 1)~.\]
The taxes for user 1 will depend on the realized exit equilibrium. If $a<N-1$, this tax is given by: 
\[t_1^P =  (N-1)\frac{c}{aN} - (N-1)\frac{c}{N-1} - c(N-1)\hat{x}^1_j =  (N-1)\frac{c}{aN} - c(1+\log\frac{N-1}{c})~. \]
The sum of the Pivotal taxes under this parameter conditions will then be given by: 
\begin{align*}
\sum_i t_i^P &= c\left(\frac{N-1}{a}\left(\log\frac{N}{N-1} - 1 + \frac{1}{N}\right) - \left(1+\log\frac{N-1}{c}\right)\right)
\end{align*}
Note that $\log{z}-\frac{1}{z}<0, \forall z<\frac32$, and therefore, with $N\geq 3$, the above sum is always negative. We conclude that the Pivotal mechanism will always carry a deficit. 

On the other hand, when $a>N-1$, the tax for the dominant user is given by: 
\[t_1^P =  (N-1)\frac{c}{aN} - (N-1)\frac{c}{a}=  (N-1)\frac{c}{a}(\frac{1}{N}-1)~.\]
The sum of the Pivotal taxes will then be given by: 
\begin{align*}
\sum_i t_i^P &= \frac{c(N-1)}{a}\left(-1+\log\frac{N}{N-1} - 1 + \frac{1}{N}\right)
\end{align*}
By the same argument as before, the above sum will always be negative, indicating a budget deficit in the Pivotal mechanism under this scenario as well.

\end{document}